\documentclass[12pt,indent]{article}
\usepackage{a4,latexsym,amsmath,amssymb}
\usepackage[latin1]{inputenc}
\usepackage{float}

\usepackage{booktabs,caption}
\usepackage[flushleft]{threeparttable}
\usepackage{rotating}

\usepackage{natbib}
\usepackage{graphics}
\usepackage[english]{babel}
\usepackage{tabularx}
\usepackage{lscape}
\usepackage{multirow}
\usepackage{rotating}
\usepackage{dsfont}
\usepackage{caption}
\usepackage{subcaption}
\usepackage{bbm}
\usepackage[table]{xcolor}
\usepackage{booktabs}
\usepackage{calc}
\usepackage{ifthen}
\usepackage{url}
\usepackage{colortbl}      
\usepackage{tikz}

\usepackage{blindtext}
\usepackage{scrextend}
\usepackage{mathtools}
\usepackage{verbatim}

\usetikzlibrary{shapes,snakes,matrix}

\newenvironment{tabularsmall}
{ \footnotesize \sffamily \tabular } {
\endtabular
\normalfont }

\newcommand{\blanco}[1]{}

\def\d{\displaystyle}

\usepackage{natbib}

\usepackage{geometry}
\usepackage{setspace}
\usepackage{tikz}
\usepackage[]{graphicx}
\usepackage[]{color}
\makeatletter
\def\maxwidth{ %
  \ifdim\Gin@nat@width>\linewidth
    \linewidth
  \else
    \Gin@nat@width
  \fi
}
\makeatother

\definecolor{fgcolor}{rgb}{0.345, 0.345, 0.345}

\usepackage{framed}
\makeatletter
 {\par\unskip\endMakeFramed%
 \at@end@of@kframe}
\makeatother

\definecolor{shadecolor}{rgb}{.97, .97, .97}
\definecolor{messagecolor}{rgb}{0, 0, 0}
\definecolor{warningcolor}{rgb}{1, 0, 1}
\definecolor{errorcolor}{rgb}{1, 0, 0}

\usepackage{alltt}
\IfFileExists{upquote.sty}{\usepackage{upquote}}{}

\begin{document}
\bibliographystyle{chicago}
\sloppy

\makeatletter
\renewcommand{\section}{\@startsection{section}{1}{\z@}%
        {-3.5ex \@plus -1ex \@minus -.2ex}%
        {1.5ex \@plus.2ex}%
        {\reset@font\large\sffamily}}
\renewcommand{\subsection}{\@startsection{subsection}{1}{\z@}%
        {-3.25ex \@plus -1ex \@minus -.2ex}%
        {1.1ex \@plus.2ex}%
        {\reset@font\normalsize\sffamily\flushleft}}
\renewcommand{\subsubsection}{\@startsection{subsubsection}{1}{\z@}%
        {-3.25ex \@plus -1ex \@minus -.2ex}%
        {1.1ex \@plus.2ex}%
        {\reset@font\normalsize\sffamily\flushleft}}
\makeatother



\newsavebox{\tempbox}
\newlength{\linelength}
\setlength{\linelength}{\linewidth-10mm} \makeatletter
\renewcommand{\@makecaption}[2]
{
  \renewcommand{\baselinestretch}{1.1} \normalsize\small
  \vspace{5mm}
  \sbox{\tempbox}{#1: #2}
  \ifthenelse{\lengthtest{\wd\tempbox>\linelength}}
  {\noindent\hspace*{4mm}\parbox{\linewidth-10mm}{\sc#1: \sl#2\par}}
  {\begin{center}\sc#1: \sl#2\par\end{center}}
}



\def\R{\mathchoice{ \hbox{${\rm I}\!{\rm R}$} }
                   { \hbox{${\rm I}\!{\rm R}$} }
                   { \hbox{$ \scriptstyle  {\rm I}\!{\rm R}$} }
                   { \hbox{$ \scriptscriptstyle  {\rm I}\!{\rm R}$} }  }

\def\N{\mathchoice{ \hbox{${\rm I}\!{\rm N}$} }
                   { \hbox{${\rm I}\!{\rm N}$} }
                   { \hbox{$ \scriptstyle  {\rm I}\!{\rm N}$} }
                   { \hbox{$ \scriptscriptstyle  {\rm I}\!{\rm N}$} }  }

\def\d{\displaystyle}\def\d{\displaystyle}

\title{Unidimensionality in Rasch Models: Efficient Item Selection and Hierarchical Clustering Methods Based on Marginal Estimates }
  \author{Gerhard Tutz \\{\small Ludwig-Maximilians-Universit\"{a}t M\"{u}nchen}\\{\small Akademiestra{\ss}e 1, 80799 M\"{u}nchen}}

\maketitle
\begin{abstract} 
\noindent
A strong tool for the selection of items that share a common trait from a set of given items is proposed. The  selection method is based on marginal estimates and exploits that the estimates of the standard deviation of the mixing distribution are rather stable if  items are from a Rasch model with a common trait. If, however, the item set is increased by adding items that do not share the latent trait the estimated standard deviations become distinctly smaller. A method is proposed that successively increases the set of items that are considered Rasch items by examining the  estimated standard  deviations of the mixing distribution. It is demonstrated that the selection procedure is on average very reliable and a criterion is proposed, which allows to identify items that should not be considered Rasch items for concrete item sets. An extension of the method allows to investigate  which groups of items might share a common trait.  The corresponding hierarchical clustering procedure is considered an exploratory tool but works well on average.  

\end{abstract}


\section{Introduction}
In item response theory it is often  assumed that the responses are determined by a single latent trait and models are based on the
assumption of unidimensionality.   Various tests, in particular for the Rasch model,  have been proposed to investigate model fit and unidimensionality.
Extensive overviews have been given by \citet{glas1995testing} and \citet{verhelst2001testing}.

Methods to investigate unidimensionality typically use test statistics as the Martin-L\"{o}f test \citep{martin1973statistiska}.
It is a likelihood ratio test for the hypothesis that the Rasch model holds
for the whole set of items against the hypothesis that this model holds for two disjoint subsets
of items and uses  conditional maximum likelihood
estimation. A major problem with approaches like that is that there is a very large number of possible disjoint subsets
of items ($2^{I-1}$, where $I$ denotes the number of items). Considering subsets that contain only one item, which would reduce the number of subsets, fails since conditional maximum likelihood needs larger item sets. In addition, the tests have to assume that the model holds within the subsets, which typically is not the case, yielding unreliable results.
Also extended versions as considered by \citet{christensen2002testing} for polytomously scored items suffer from these problems. Moreover, as \citet{bartolucci2007class} notes, the resulting test statistic has not, under the null hypothesis, an easy-to-handle asymptotic distribution. He also describes  problems with alternative testing approaches arising for tests proposed by \citet{van1982two}, \citet{glas1991contributions}. 

The approach proposed here does not rely on test statistics to avoid the problems that come with the methods. 
Instead we exploit that  marginal estimates yield stable estimates of the variance of the mixing distribution if the Rasch model holds but yield small  values if items that do not share the latent trait are included. In general, selection methods should be tailored to the purpose. We consider in particular two scenarios. In the first a fairly homogeneous 
set of items is assumed to be given but it is unclear if all the items share the  trait to be measured, one suspects that some of the items do not follow the   model. The method proposed for this scenario can also be seen as a forward selection procedure aiming at selecting items that are  compatible with the Rasch model. An advantage of the method is that it avoids  problems that arise with traditional selection methods. A widely used method is based on 
Andersen`s likelihood ratio test \citep{andersen1973goodness}, which exploits that  item parameters can be  consistently estimated in any sample of persons provided the Rasch model holds in the corresponding population. Comparing  the estimates obtained in defined subgroups of subjects can be used to detect which items that do not fit well. The problem  is that the results depend on the choice of the subgroups. Different subgroups will yield different items as problem items and often there  is a plethora of subsets that could be considered.

In the second scenario the given set of items is assumed to be more  diverse, items might come from quite different Rasch models. For this scenario a hierarchical 
clustering procedure  that separates the different item sets is proposed. There have been several attempts to construct hierarchical clustering procedures, for example the testing approach propagated by \citet{bartolucci2007class}. He developed a general framework based on a class of multidimensional IRT models and uses test statistics to investigate unidimensionality, and also proposes a
hierarchical clustering of the items, so that the items in the same group are referred to the same latent trait.
Alternative cluster methods with a focus on nonparametric
item response theory have been reviewed by \citet{van2004comparative}. The selection of items from multi-dimensional datasets  in the context of the Mokken item response theory model have been propagated by \citet{hemker1995selection}. The method is based on the  scalability coefficient, which uses the covariance between item scores. Other methods that use conditional covariances have been considered by \citet{zhang1999theoretical}, \citet{zhang1999conditional}.

It is crucial that  cluster methods that rely on association measures or distances can not be expected to work well. One way to define association measures is to consider correlations or covariances, which can be transformed to distance measures by using a strictly decreasing transformation. However, correlations between items are very small  yielding poor cluster results. To illustrate the point Table \ref{tab:corr1} (left hand side) shows exemplarily the estimated correlations between items for one data set in which the Rasch model holds (200 persons). 
It is seen that correlations are definitely small, the correlation between item 1 and item 4 is just 0.02. In some simulations also negative correlations can be observed, although the Rasch model holds. Thus, deriving distances between items from correlations is not very promising for Rasch models although they might be useful in nonparametric approaches as considered by \citet{hemker1995selection}. Also  distances between items as the Euclidean distance between response vectors do not work well.  

An alternative to correlations are conditional correlations or covariances as propagated by \citet{zhang1999theoretical}. The criterion they use to judge if two items are substantially similar  is that the conditional covariance of the item pair given the sum score is positive. Although the theoretical reasoning is  convincing it is not so easy to estimate the conditional covariance. The sample covariance given the sum score, which might be used to estimate the conditional covariance given the Rasch model holds, is typically negative although the Rasch model holds. Table \ref{tab:corr1} (right hand side) shows the estimated conditional covariances (averaged over all sum scores) for the same data set for which the correlations have been computed. Although values are close to zero  they are negative, not positive. In addition, conditional correlations can not be computed for all sum scores since empirical variances are zero for some sum scores.

\begin{table}[H]
 \caption{Correlation and conditional covariance for four Rasch item } \label{tab:corr1}
\centering
\begin{tabularsmall}{llllllllcccccccccc}
  \toprule
 &\multicolumn{5}{c}{Correlation } &\multicolumn{5}{c}{ Conditional covariance } \\
 &&Item1&Item2&Item3&Item4&&Item1&Item2&Item3&Item4\\
 &Item1 &1.00  &0.18 &0.22 &0.02 & &0.154 &-0.018 &-0.028 &-0.058\\
 &Item2&0.18 &1.00 &0.13 &0.14 &  &-0.018  &0.108 &-0.036 &-0.022\\
 &Item3&0.22 &0.13 &1.00 &0.10 & &-0.028 &-0.036  &0.150 &-0.040\\
 &Item4&0.02 &0.14 &0.10 &1.00 &  &-0.058 &-0.022 &-0.040  &0.174\\
 \bottomrule
\end{tabularsmall}
\end{table}

\section{Marginal Estimates for the Rasch Model}  \label{sec:marg}

Let   persons or test takers $1,\dots, P$ be confronted with a set of similar tasks or items $1,\dots, I$ to be mastered.  The response $Y_{pi}$ of person $p$ on item $i$ takes values from $\{0,1\}$ with '1' indicating success and '0' indicating failure. 
If the Rasch model holds the probability of success is given by 
\begin{equation}\label{eq:BinRasch}
P(Y_{pi}=1|\theta_p,\delta_{i})=\frac{\exp(\theta_p-\delta_{i})}{1+\exp(\theta_p-\delta_{i})},
\end{equation} 
where 
 $\theta_p$ is the ability of person $p$, and
 $\delta_{i}$ is the difficulty of item $i$.

There are several ways to obtain estimates. One option is to estimate item parameters by using marginal log-likelihood.
Marginal maximum likelihood (MML) estimates of item parameters are obtained by assuming a fixed distribution of person parameters, a typical choice is the normal distribution. The corresponding marginal likelihood has the form
\begin{align*}
L_m &= \prod_{p=1}^P \int \prod_{i=1}^I P( Y_{pi}=1)^{y_{pi}}P( Y_{pi}=0)^{1-y_{pi}}  f_{\sigma_{\theta}}(\theta_p) d\theta_p,
\end{align*}
where $f_{\sigma_{\theta}}(\theta_p)$ is the density of a  normal distribution with mean zero and standard deviation $\sigma_{\theta}$. Also alternative distributions can be used, see, for example, \citet{andersen1977estimating}, \citet{thissen1982marginal}.  
The marginal estimates based on the normal distribution is available in several program packages, a very efficient one that is used in the following is TAM 
\citep{robitzsch2022package}.

\begin{figure}[H]
\centering
\includegraphics[width=7cm]{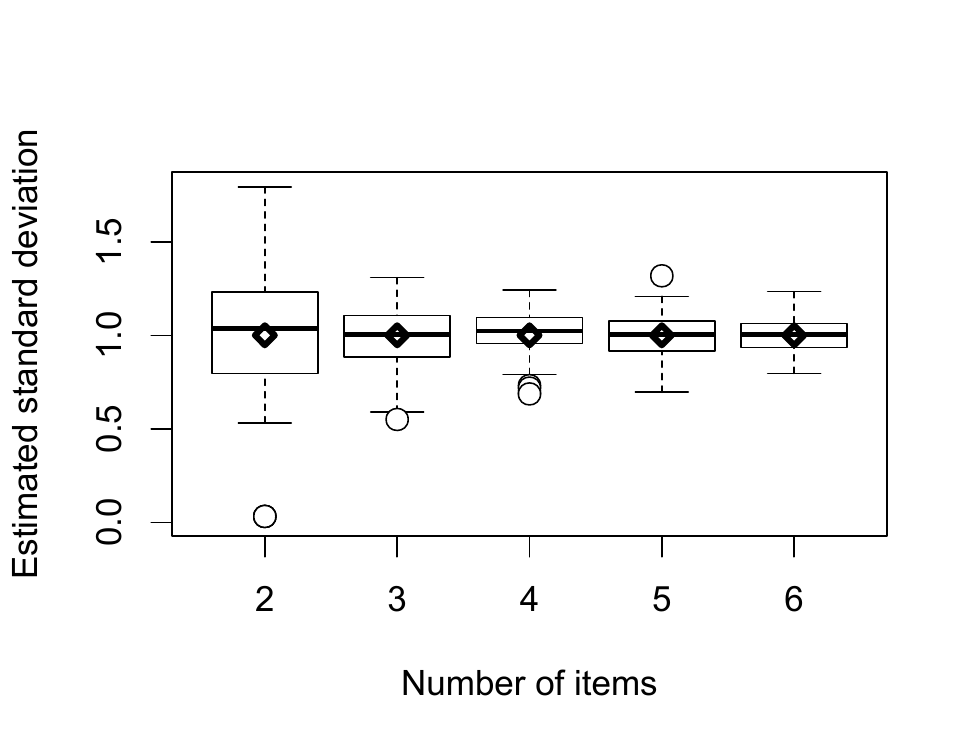}
\includegraphics[width=7cm]{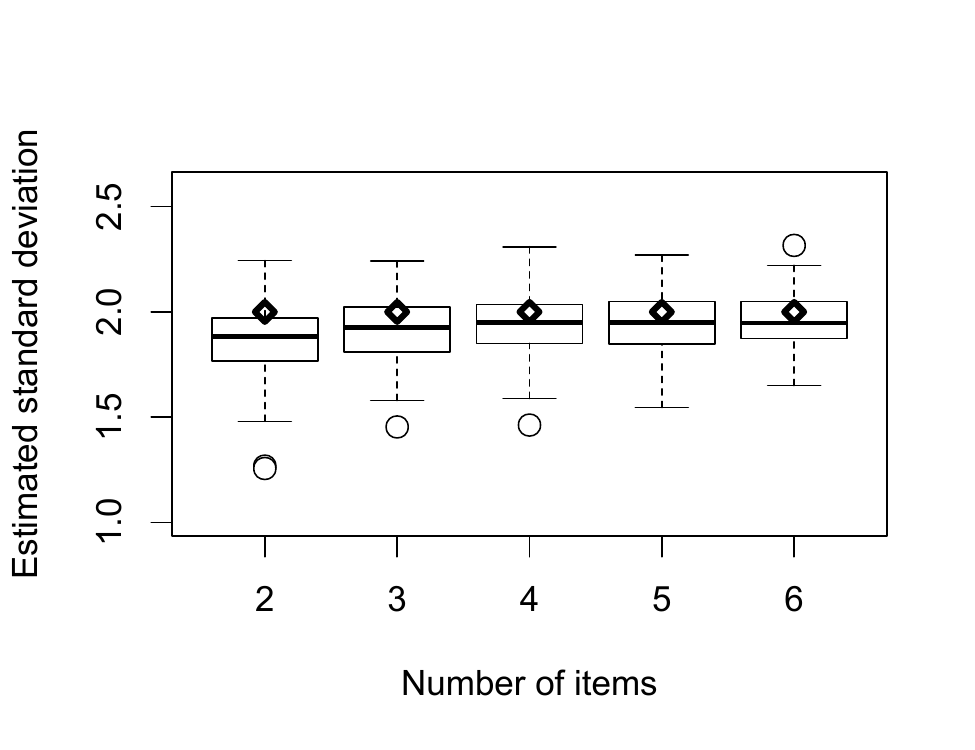}
\includegraphics[width=7cm]{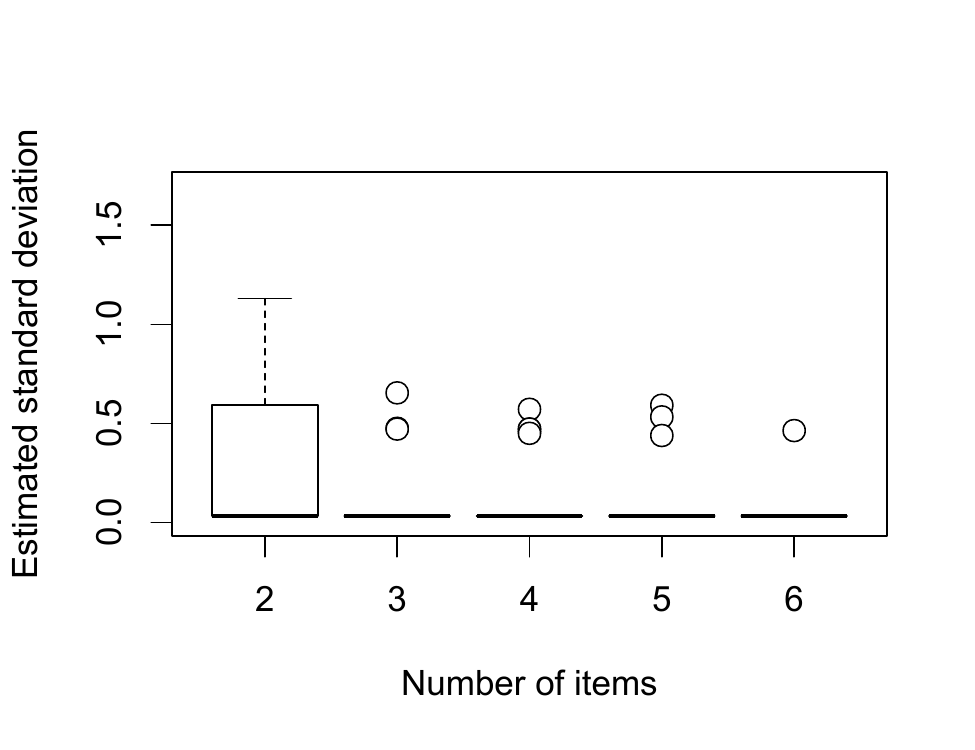}
\includegraphics[width=7cm]{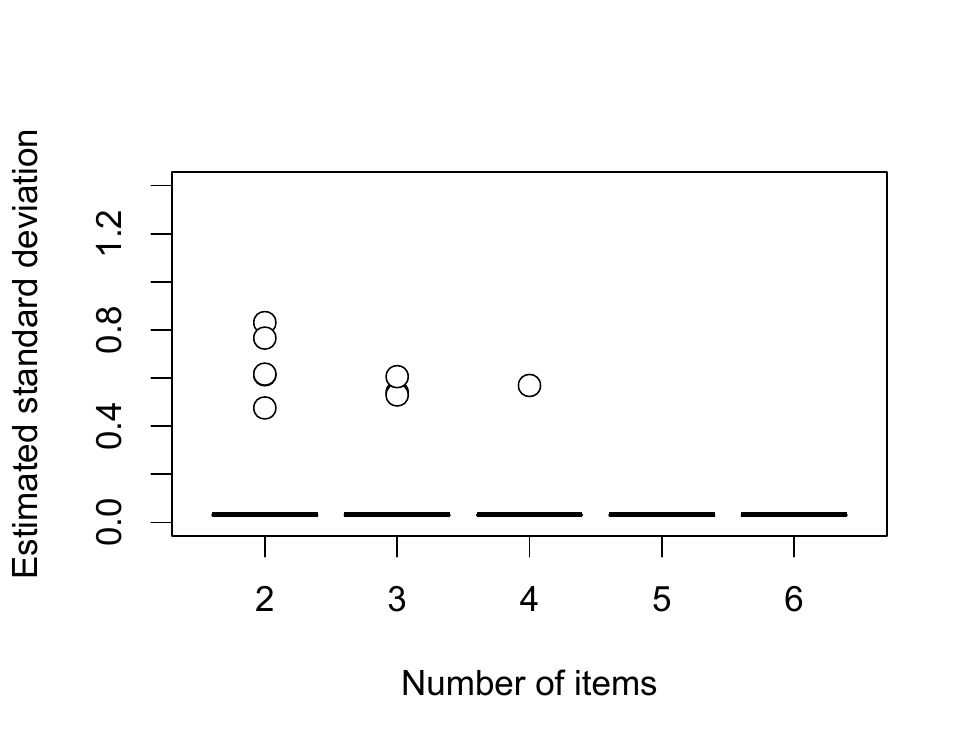}
\caption{Standard deviation for 6 items, first row: generated from Rasch model, second row: independent traits, first column: standard deviation is 1, second column: standard deviation is 2}
\label{fig:var}
\end{figure}

Typically one is interested in the estimation of the difficulty parameters parameter and the estimation of the standard deviation of the mixing distribution is of minor interest. However, the standard deviation of the mixing distribution, $\sigma_{\theta}$, can be very useful in the investigation of unidimensionalty.
If items following a Rasch model are of any use the standard deviation $\sigma_{\theta}$ has to be distinctly larger than zero, therefore in the following it is assumed that the standard deviation is definitely larger than zero. 
 Figure \ref{fig:var} shows the estimates for 6 items that follow a Rasch model with difficulties 0, -1.5, -1,   0.5, 1.2, 1.5 
and 200 persons. The  box plots show the estimates of the standard deviations if increasing numbers of items are included. The first box plot shows the estimates if only   the first two items are used, the second if the first three items are used, up to all 6 items. The left picture shows the estimates if the true standard deviation is 1, the right picture for 
$\sigma_{\theta}=2$ (given as diamonds). It is seen that the standard deviation is estimated rather well, only if the number of items is very small and the true standard deviation  is comparable small the variance of the estiamtes is rather large. The crucial point that is used in the following is that the estimated standard deviation tends to be very small if the items represent different traits. The second row of Figure \ref{fig:var} shows estimates that are obtained after the responses for each item have been randomly permuted, therefore representing independent traits. Again the first box plot shows only the first two items, the second the first three, up to all 6 items. The left picture shows the estimates if the true standard deviation is 1, the right picture for 
$\sigma_{\theta}=2$. It is seen that the estimates tend to be very small indicating that the items do not share a  common trait.
It shows that the (estimated) standard deviation of the mixing distribution might be used as an indicator of the homogeneity of items, that is, their compatibility with a Rasch model.

For further illustration let us consider again six items, the first five items follow a Rasch model with a one-dimensional trait while item 6 does not. It is drawn from a Rasch model but the responses have been permuted randomly. Table \ref{tab:respt1} shows the estimated standard deviations if the Rasch model is fitted in clusters of two items (item parameters as in the previous simulation, $\sigma_{\theta}=1$). For example, if one fits the model in cluster $C_{12}=\{1,2\}$, which contains only items 1 and 2, the estimated standard deviation is 1.002. It is seen that the standard deviations are large if clusters are built from the set $\{1,\dots,5\}$, for which the Rasch model holds,  but are always very small   if item 6 is involved (not all possible clusters shown).
The case of just two items is the most critical one since estimates are less stable (as seen from Figure \ref{fig:var}) if only few items are used and the standard deviation is comparatively small ($\sigma_{\theta}=1$). Nevertheless, pairs of items for which the Rasch model holds yield much larger estimates  
than pairs that include item 6.

\begin{table}[H]
 \caption{Estimated standard deviation for various clusters that contain two items (six items available, item 6 is not a Rasch item )} \label{tab:respt1}
\centering
\begin{tabularsmall}{llllllllcccccccccc}
  \toprule
 &   & item1 & item2 &   estimated std deviation\\    
  \midrule
 &$C_{12}$    &1    &2    &      1.002\\
 &$C_{13}$   &1    &3    &      0.732\\
  &$C_{16}$    &1    &6    &      0.032\\
  &$C_{24}$    &2    &4    &      1.286\\
 &$C_{34}$    &3    &4    &      0.706\\
&$C_{26}$    &2    &6    &      0.032\\
&$C_{56}$    &5    &6    &        0.032 \\
\bottomrule
\end{tabularsmall}
\end{table}

\section{Item purification: identifying non-Rasch items }\label{sec:id}

First we will consider strategies for the case where one has a fairly homogeneous set of items but 
it is unclear if \textit{all} the items are compatible with the model. The objective is to exclude those items that do not follow the Rasch model.

The strategy  to identify ``non-Rasch'' items  is to form clusters of items, more precisely, one big cluster that contains the Rasch items for the trait that is to be measured and several clusters that contain items for which responses are not generated by the same Rasch model. The items in the latter clusters might be Rasch items but for   different traits. 

As tool in cluster building we use the estimated standard deviation of newly built clusters when clusters are fused. Let $P=(C_1,\dots,C_m)$ be a 
disjunct partitioning of the items with $C_i \cap C_j=\emptyset$ for $i \ne j$ and $C_1 \cup \dots C_m =\{1,\dots,I\}$.  Let $P_{ij}$ denote the partition that is built from $P$ by fusing the clusters $C_i, C_j$ from partition $P$, the others are identical to clusters from $P$. The indicator of compatibility with the Rasch model is the size of the estimated standard deviation of the newly built cluster, which results from fusing clusters $C_i, C_j$, 
\begin{equation}\label{equ:hetf}
S(P,P_{ij})= \sigma_{\theta_p}(C_i \cup C_j),
\end{equation}
where $\sigma_{\theta_p}(C_i \cup C_j)$ is the estimated standard deviation of marginal estimates if Rasch models are fitted for items $C_i \cup C_j$.
$S(P,P_{ij})$ can be seen as a homogeneity measure. Large values, in the range of the true standard deviation indicate that the items in the cluster $C_i \cup C_j$ 
are Rasch items with a common trait, small values indicate that at least some of the items in the combined cluster are non-Rasch items.

\vspace{0.5cm}
\hrule
\begin{center}{\textbf{Clustering I: Finding Rasch items that share a common trait}}\end{center}

\begin{description}
\item{\textit{Step 1 (Initialization)}}

\begin{itemize}
\item[(a)] Computation

Let $P_0=(C_1,\dots,C_I)$, $C_j=\{j\}$ be the starting partition, in which each cluster is formed by one  item.   Let  $P_{ij}$ denote the partition that contains the $I-1$ clusters $C_i \cup C_j$, $C_s, s\ne i,j$, which is built by fusing $C_i$ and $C_j$. Let $s_{ij}=\sigma_{\theta_p}(P_0,P_{ij})$ denote the standard deviation of if $P_{ij}$ is built from $P_0$.
\item[(b)] Selection

Select the partition with the largest fusion homogeneity $s_{ij}=\sigma_{\theta_p}(C_i \cup C_j)$,
\[
(i_0,j_0) = \operatorname{argmax}_{(i,j)} s_{ij}.
\]
\item[(b)] Cluster fusion

Form the new partition $P_1$ by fusing the clusters $C_{i_0}, C_{j_0}$ yielding
$P_1=(C_1^{(1)},\dots,C_{I-1}^{(1)})$ with $C^{(1)}_{1}=\{i_0,j_0\}$ and  the other clusters being single item clusters that contain items $s, s \ne i_0,j_0$.
\end{itemize}

\item{\textit {Step 2 (Iteration)}}
For $\ell=1,\dots,I-2$
\begin{itemize}
\item[(a)] Computation  

Let  $P_{\ell j}$ denote the partition that contains the  clusters $C_1^{(\ell)}\cup C_j^{(\ell)}$, $j=2,\dots,I-\ell$, 
which is built by a fusion of $C_1^{(\ell)}$ and $C_j^{(\ell)}$. 
Compute the  corresponding standard deviations   $s_{\ell j}=\sigma_{\theta_p}(P_{\ell},P_{\ell j})$.
\item[(b)] Selection

Select the partition that yields the largest standard deviation $s_{\ell j}$ when fusing two clusters,
\[
j_0 = \operatorname{argmax}_{j} s_{\ell j},
\]
\item[(b)] Cluster fusion

Form the new partition $P_{\ell+1}$ by fusing the clusters $C_1^{(\ell)}$, $C_{j_0}^{(\ell)}$ to obtain
$P_{\ell+1}=(C_1^{(\ell+1)},\dots,C_{I-\ell-1}^{(\ell+1)})$ with $C_1^{(\ell+1)}=C_1^{(\ell)} \cup C_{j_0}^{(\ell)}$ and 
$C_j^{(\ell+1)}=C_j^{(\ell)}$, $j \ne 1,j_0$.
\end{itemize}
\hrule 
\vspace{0.5cm}
\end{description}

In the initialization step a pair of items is selected that shows the largest standard deviation. It forms the first cluster. In the following steps this set of items is enlarged. 
Apart from the initialization step it is a forward selection strategy. The algorithm can be easily modified such that it starts with a fixed anchor item, which is known to be a Rasch item for the latent trait  to be measured. Then the initialization step is like the iteration steps, it considers only pairs of clusters that contain the fixed Rasch item. However, it assumes that an anchor item  item is known. In the following we do not consider this version because it assumes that more information is available, and, as will be demonstrated, the version without this knowledge  performs rather well.

The algorithm yields a sequence of clusters $C_1^{(1)},\dots,C_1^{(I-1)}$ derived from the partitions $P_1,\dots,P_{I-1}$.
The sequence of clusters represent the sets of items that are considered Rasch items at specific steps of the algorithm. Cluster $C_1^{(j)}$ contains $j+1$ items, the last one, $C_1^{(I-1)}$ , contains all items. 

The stability of the estimated standard deviation over steps is an indicator of the homogeneity of items. If items that are included share the same trait as 
the previously included items the standard deviation is expected to remain in the range of the underlying true standard deviation. However, if non-Rasch items are included, that is, items that do not share the same trait, the standard deviation will decresase.

\begin{figure}[h!]
\centering
\includegraphics[width=7cm]{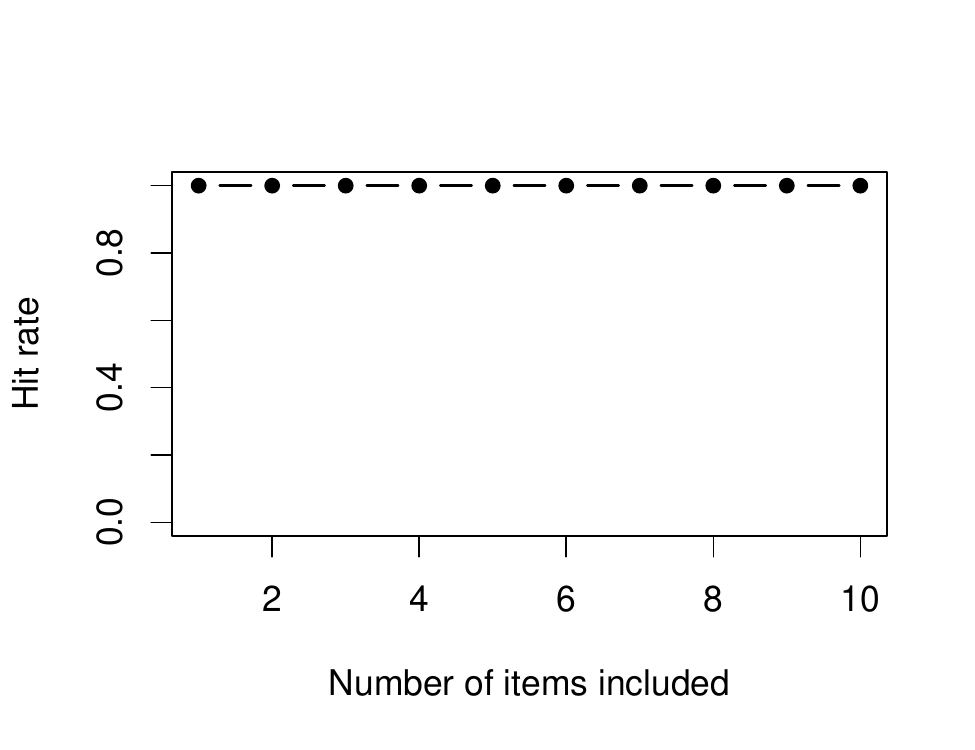}
\includegraphics[width=7cm]{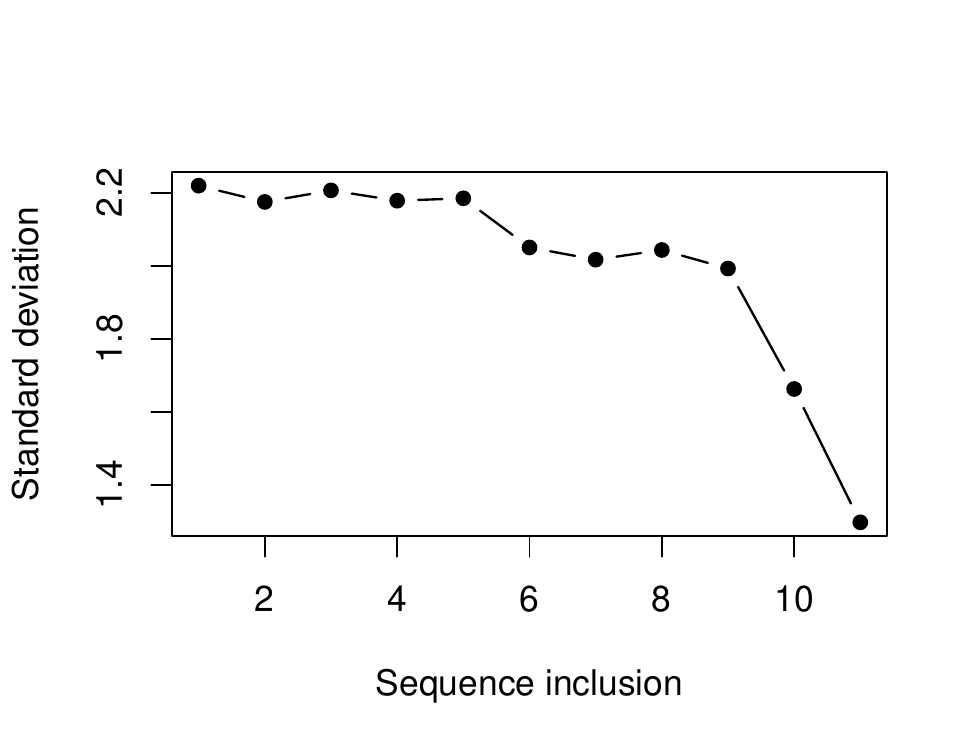}
\includegraphics[width=7cm]{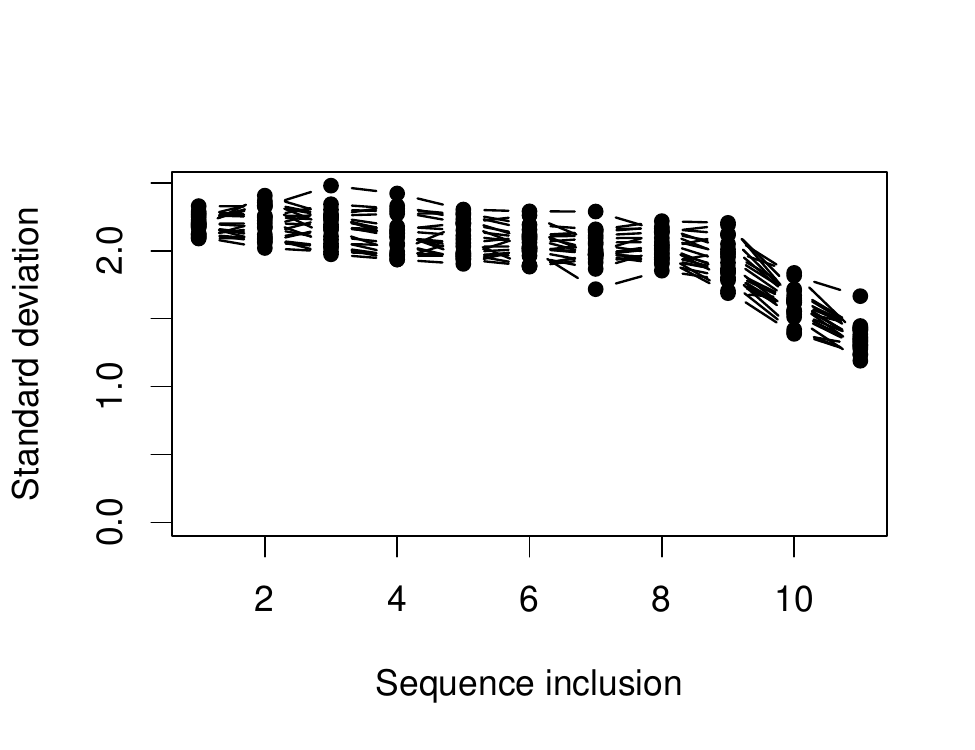}
\caption{Hit rate, standard deviation for 12 items, items 11 and 12 are non-Rasch, $P=200$, $\sigma_{\theta}=2$}
\label{fig:hettw2}
\end{figure}

\begin{figure}[h!]
\centering
\includegraphics[width=7cm]{hit12-3}
\includegraphics[width=7cm]{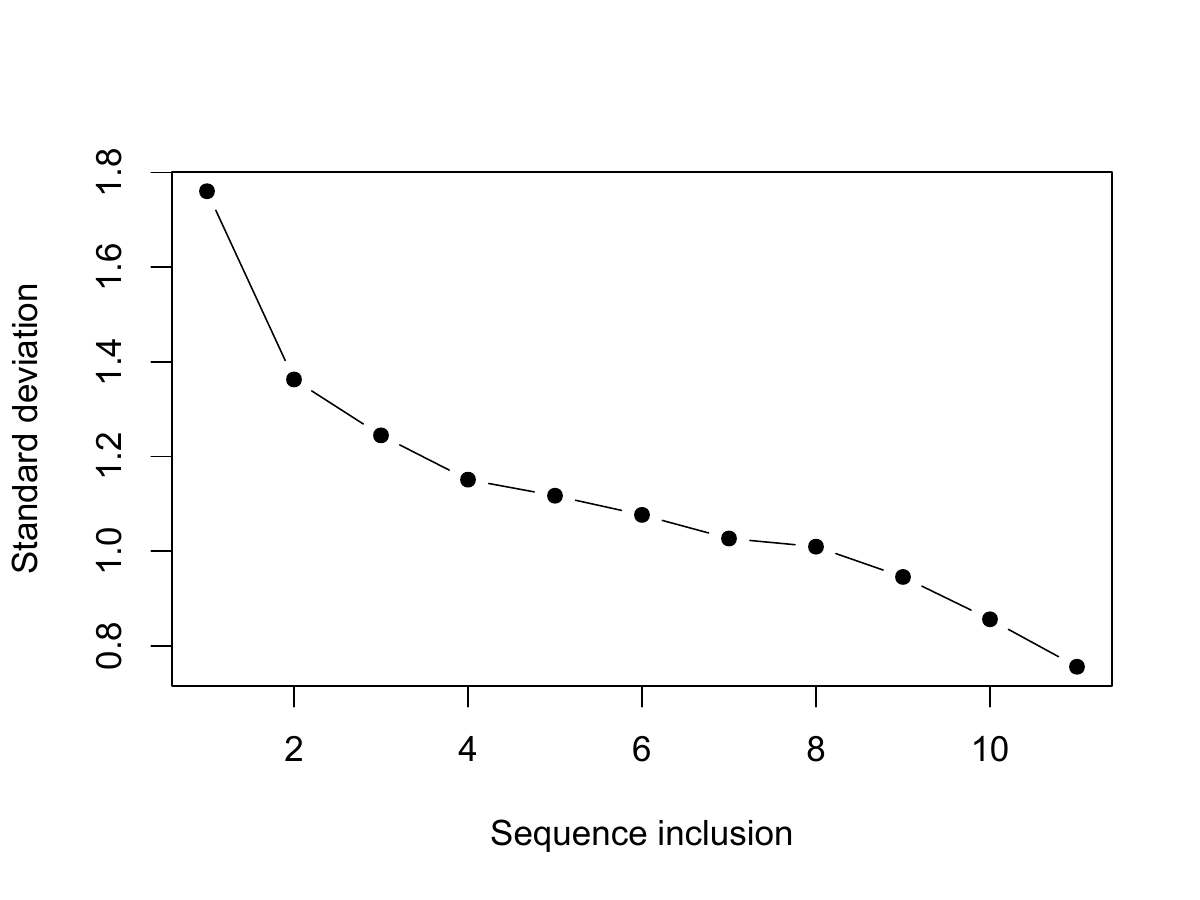}
\includegraphics[width=7cm]{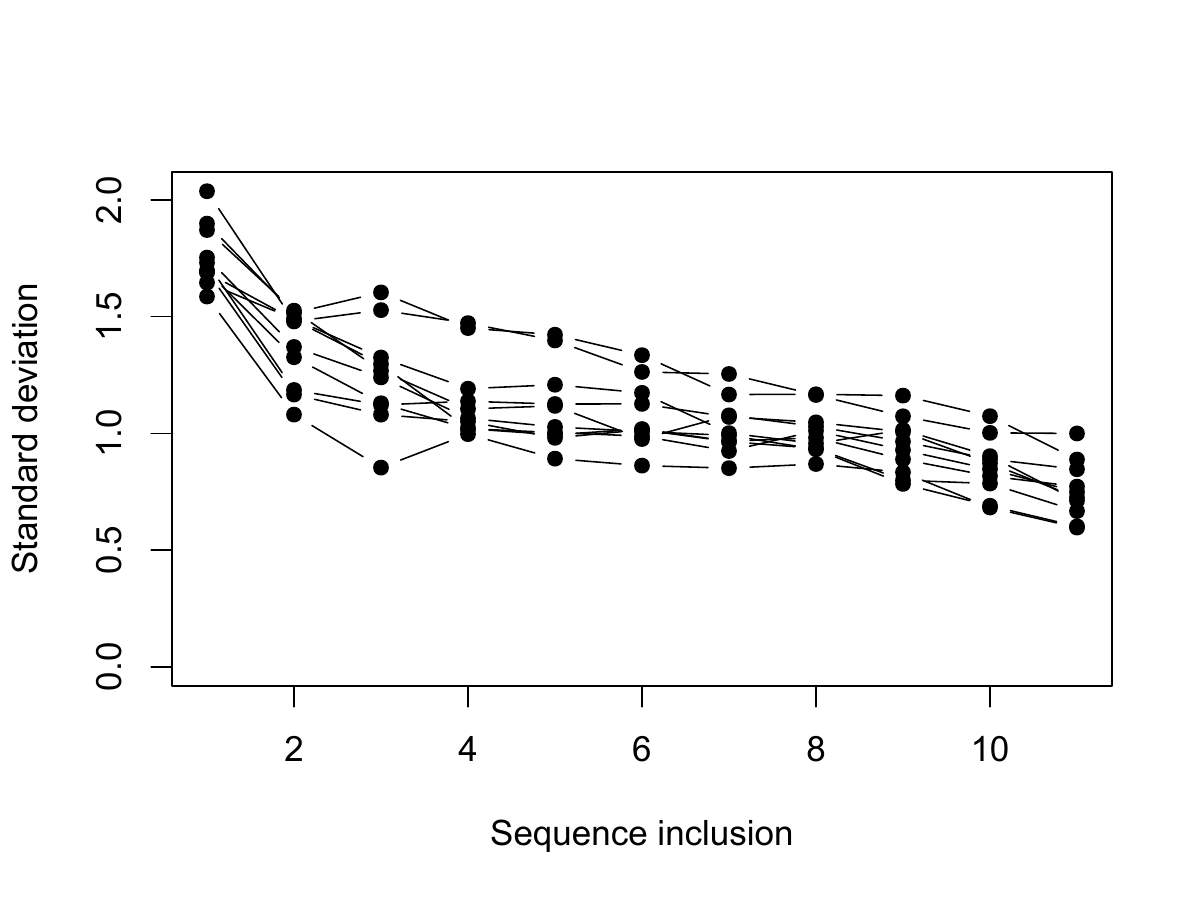}
\caption{Hit rate, standard deviation for 12 items, items 11 and 12 are non-Rasch, $P=200$, $\sigma_{\theta}=1$}
\label{fig:hettw1}
\end{figure}

For illustration Figure \ref{fig:hettw2} shows what happens if responses are generated by a Rasch model with 12 items with difficulties 0.0 -1.5 -1.0  0.5  1.2  1.5  0.2 -1.3 -0.8  0.7  1.4  1.7 ($\sigma_{\theta}=2$). Items 1 to 10 are Rasch items but items 11 and 12 are not. The algorithm was run for 50 data sets.
The first picture in Figure \ref{fig:hettw2} shows the hit rate, that is, the proportion  of the simulation runs in which  an item has been selected that is from the set of Rasch items. It is seen that selection is perfect, up to 10 items that are included none was a non-Rasch item.
The second picture shows the estimated standard deviation for the fused clusters, that is, for the clusters that have been selected for fusion  as a function of the inclusion steps. In the first step two items are included, in the second three items are in the selected set, etc. It is seen that the standard deviation remains rather stable in the first nine steps but distinctly drops in step 10 of the algorithm. In this step for the first time one of the non-Rasch items was included. The third picture shows the standard deviation as a function of the inclusion steps for the first 10 simulated data sets. The distinct decrease in step 10 is found in all of the 10 data sets. It suggests that  the decrease can taken as an indicator that a non-Rasch item is included. 
However, the drop in standard deviation is less pronounced if the true standard deviation is smaller as is seen from Figure \ref{fig:hettw1}, which shows the same pictures but for true standard deviation $\sigma_{\theta}=1$. Although the hit rates  still show perfect selection of Rasch items and the standard deviation decreases in step 10, there is already a strong decrease within the first steps. Therefore the decrease itself is not a reliable indicator of the inclusion of wrong items. 
The reason is that standard errors of the estimation are larger if only few items are included and the standard deviation is small. It should also be mentioned that the estimates should not be seen as good estimates of the true standard deviation. Since the clusters that show maximal value are selected the estimates tend  to overestimate the standard deviation when clusters of items are considered for which the model holds. However, this does not affect the selection of items, which is based on \textit{comparing} standard deviations. Not the estimated value, which is biased, is important but the comparison of estimates. Nevertheless, the decrease in standard deviation is not a good indicator that non-Rasch items are included, therefore a stronger criterion is needed to determine when to stop the inclusion of items, to be considered in the next section.

\subsection{Selection Stability and misfit scoring  }

Item selection by building clusters works very well and items that do not follow the Rasch model are included in the last steps only. In applications items that are selected very late should be considered more closely from a content perspective. Are  the  items actually measuring the same trait as the other items or are they constructed quite differently from the other items. 

However, in applications one obtains a single sequence of items that are included successively as Rasch items into the first cluster and  a criterion is needed that describes when the items included should be considered non-Rasch items. As is seen from Figures  \ref{fig:hettw2} and  \ref{fig:hettw1}
the inclusion of wrong items is accompanied by an decrease   of estimated standard deviations. If they decrease strongly it might be taken as an indicator that suspicious items are included. The problem  is that sometimes decreases are already seen in early steps of the algorithm when true Rasch items are included.  

The stopping criterion developed in the following focuses on the stability of the selection of items in subsets. If an item is not a Rasch item it should be included very late in \textit{all} subsets of persons.
Therefore, we investigate  randomly drawn  subsets of persons $S_1,\dots S_M$ and run the algorithm for each subset to obtain  $M$ sequences of items  
\[
i_1^{(m)},\dots, i_I^{(m)}, \quad  m=1,\dots, M,
\]
where $i_1^{(m)}$ is the first item that is included when using subset $S_m$, $i_2^{(m)}$ is the second item that is included, etc.
Let 
\[
o_{im} = \operatorname{arg}_j i_j{(m)}
\]
denote the order of item $i$ in the sequence of items $i_1^{(m)},\dots, i_I^{(m)}$ when using subset $S_m$.
This yields for each item $M$ observations of inclusion order collected in the set 
$O_i= \{o_{i1},\dots,o_{iM} \}$.
The distribution of the values in $O_i$ is used as an indicator of the compatibility of item $i$ with the Rasch model. 

\begin{figure}[h!]
\centering
\includegraphics[width=7cm]{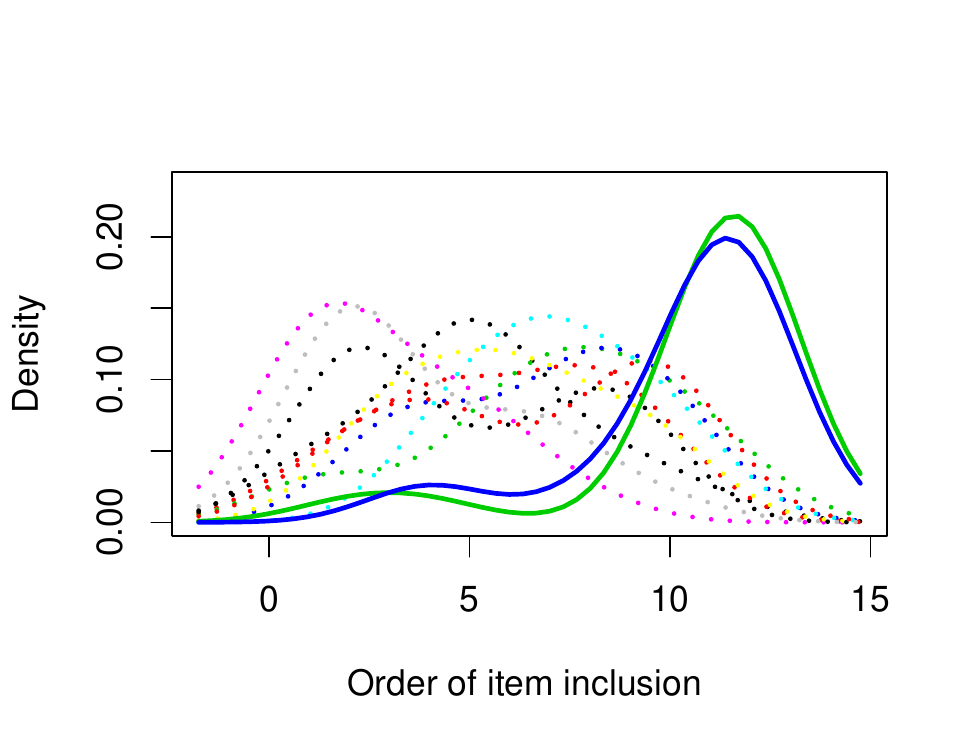}
\includegraphics[width=7cm]{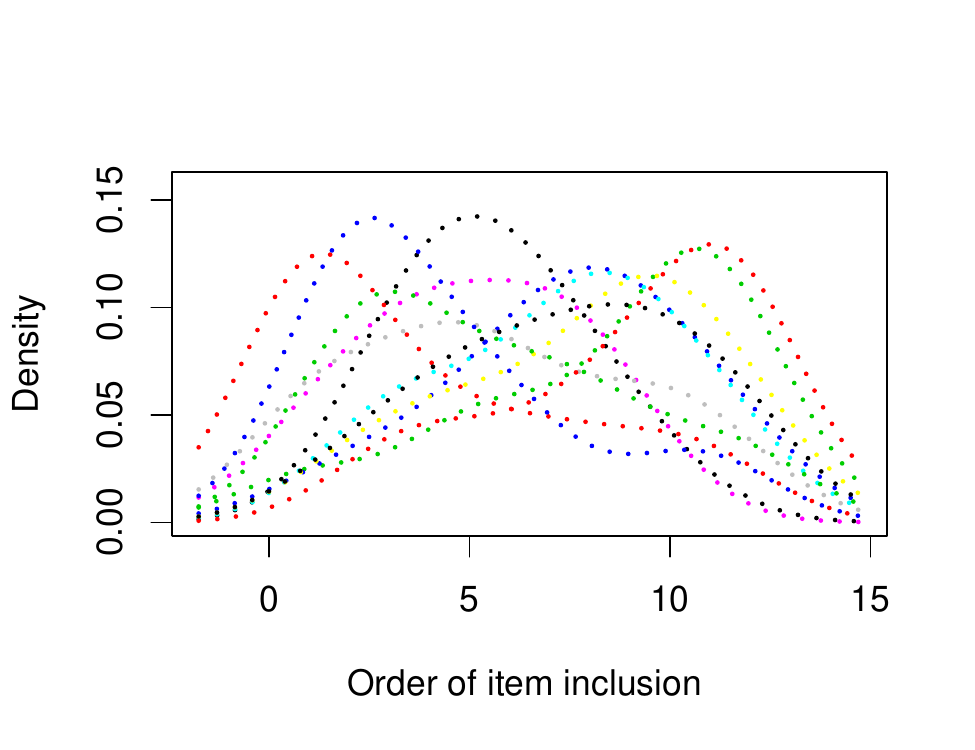}
\caption{Distribution of order of inclusion for selected items, 12 items,  4 sequences; left picture: items 11,12 are non-Rasch items, right picture: all items are Rasch items.  }
\label{fig:math2}
\end{figure}

Figure \ref{fig:math2}  shows  typical distributions for the case of 12 items when items 11,12 are non-Rasch items (left picture, kernel smoothed estimates). The proportion of persons in subsamples was 0.5 (also used later).   Items 11 and 12 are depicted by drawn lines, items 1 to 10 by dotted lines. It is seen that items 11 and 12  are included almost exclusively in the last steps of the algorithm. 
The right picture in Figure \ref{fig:math2} shows  typical distributions obtained when all items are Rasch items (20 subsets, all items given by dotted lines). It is seen that in the case of polluted items (11,12 are not Rasch items) the distributions of the non-Rasch items are centered at large values while in the case in which all items are Rasch items no single item stands out.  

One way to use this strategy is to look at the distributions, which show which items tend to have large values. But also a more formal criterion
can be derived from the distributions. We propose to investigate how often an item is found late in the sequences of inclusion. Let
\[
\operatorname{mf}_i = \frac{|\{o_{im}|o_{im} > 0.75 \times I, m=1,\dots,M\}|}{M},
\]
denote the number of subsets in which item $i$ is included in the last quarter of the sequence of included items divided by the number of subsets.
$\operatorname{mf}_i$ is a misfit score, it is an  indicator of the misfit of item $i$ if a Rasch model is assumed to hold. A large  value (close  to one) indicates that the item is not a Rasch item. Of course one can use a different threshold $a$ and consider items with 
$\operatorname{Ind}_i >a$ as items that should be investigated more closely, however,  $a=0.75$  works well and is used throughout the paper.  
Also other criteria could be used as misfit scores, for example, the mean  of the item order.

Table \ref{tab:ord1} shows the misfit scores for the items for which the distributions are given in Figure \ref{fig:math2}. It is seen that  
in the case where all items are Rasch items none of the misfit scores is larger than 0.5. In the case where items 11 and 12 are non-Rasch items these two items have values 0.90 and 0.80, indicating that the items are not from the same model.  In the lower part the misfit scores are shown for an average over 50 data sets. It is again seen that there are no outstanding values if the Rasch model holds but if items 11 and 12 are non-Rasch items the values are quite large. Table \ref{tab:ord2} shows the averaged mis-fit indicator values for a case with larger mixing standard deviation ($\sigma_{\theta}=2$). Then the results are even more distinct. For further simulations see appendix.

The investigation of subsamples should not be seen as a stopping criterion only. It yields misfit scores for the compatibility of items  with the Rasch model. They are more reliable than the order of an item in a single selection run and should always be computed if items are to be selected.  
The resulting misfit scores were inspired by stability selection as considered by \citet{meinshausen2010stability}. They  developed a general framework for stable variable selection (not item selection)  by subsampling.

\begin{table}[h!]
 \caption{Misfit scores for items (12 items, 20 subsets, proportion 0.5, $\sigma_{\theta}=1$)} \label{tab:ord1}
\centering
\begin{tabularsmall}{llrrrrrrrcccccccccc}
  \toprule
   item&1 &2 &3 &4 &5 &6 &7 &8 &9 &10 &11 &12\\ 
  \midrule
Single data set\\
\midrule
Rasch items      & 0.05 &0.15 &0.55 &0.25 &0.25    &0.00 &0.35  &0.20  &0.30   &0.60  &0.15  &0.15\\
Polluted case   &0.05  &0.10 &0.25  &0.20  &0.20    &0.00 &0.15 &0.05    &0.00   &0.30   &\textbf{0.90}   &\textbf{0.80}\\
\midrule
Averaged \\
Rasch items      &0.18 &0.16 &0.12 &0.36 &0.12 &0.24 &0.38 &0.40 &0.28 &0.24 &0.42 &0.10 \\
Polluted case   &0.08 &0.14 &0.16 &0.18 &0.12 &0.12 &0.20 &0.08 &0.17 &0.18 &\textbf{0.78}   &\textbf{0.80}\\
\midrule

\bottomrule
\end{tabularsmall}
\end{table}

\begin{table}[h!]
 \caption{Misfit scores for items (12 items, 20 subsets, proportion 0.5, $\sigma_{\theta}=3$)} \label{tab:ord2}
\centering
\begin{tabularsmall}{llrrrrrrrcccccccccc}
  \toprule
   item&1 &2 &3 &4 &5 &6 &7 &8 &9 &10 &11 &12\\ 
  \midrule

Averaged \\
\midrule
Rasch items      &0.31 &0.31 &0.22 &0.32 &0.17 &0.18 &0.29 &0.21 &0.35 &0.24 &0.23 &0.17\\
Polluted case   &0.14 &0.18 &0.09 &0.12 &0.06 &0.03 &0.06 &0.08 &0.11 &0.14 &\textbf{1.00} &\textbf{1.00}\\

\bottomrule
\end{tabularsmall}
\end{table}

\subsection{Illustrating data sets}\label{sec:ill}

In the following first the stability of selection is investigated for an  item set that is assumed to measure a uni-dimensional trait. Then a data set is considered for which the assumption is less founded and problems with conventional item selection methods are described. 

\subsubsection*{FIMS data}

A  FIMS dataset  study available from the TAM package \citep{robitzsch2022package} contains responses on 14 mathematics items. We use the data collected for Australian students (data.fims.Aus.Jpn.scored). Having been used as part of a FIMS study it can be expected that the items share a common trait, namely ability in mathematics.
The selection procedure proposed here is used to investigate if this claim is warranted.

\begin{figure}[h!]
\centering
\includegraphics[width=7cm]{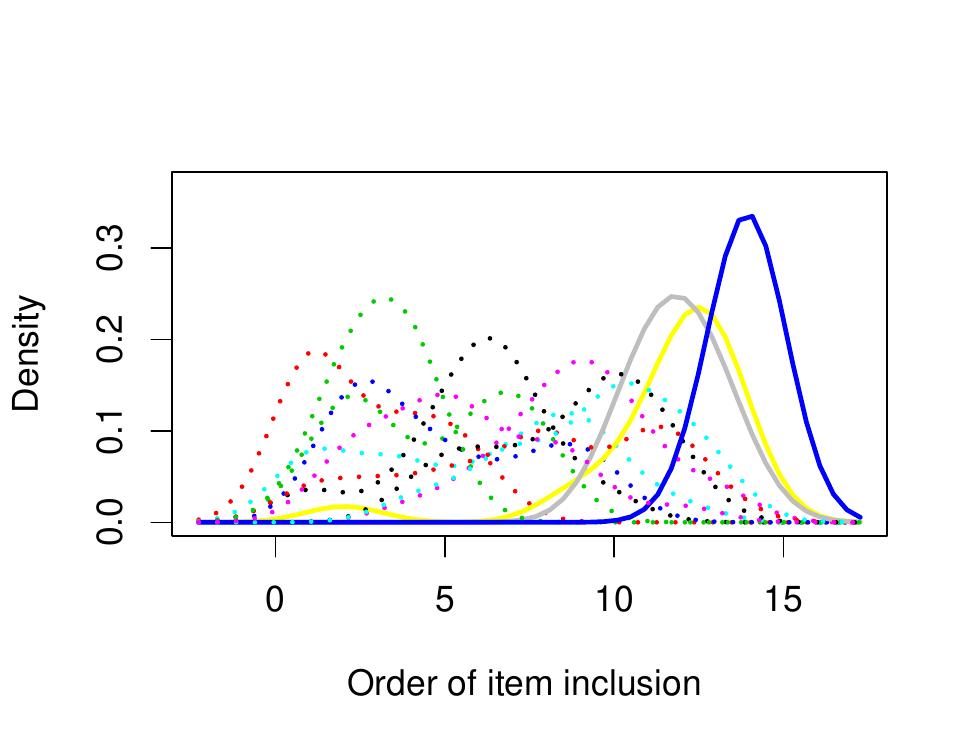}
\caption{Distribution of steps in which items are included for selected items (FIMS data)}
\label{fig:math1}
\end{figure}

Figure \ref{fig:math1} shows the distributions of the order of item inclusion. It is seen that three items, item 7,8 and 12 (shown by drawn lines) are always included very late, all other items (dotted lines) show a far less extreme  distribution. Using the citerion that  the proportion of data sets in which an item is included in the last quarter of steps  one obtains 1 for the most extreme item  12. For items 7 and 8 one obtains  0.85 and 0.9, respectively, which shows that they are also included very late. That makes them  suspicious items, which might need  more attention if the item set is used in the future.   
Table \ref{tab:fimsind} shows the misfit scores  for all items. The values for items 7,8 and 12 are given in boldface.
In addition the mean of the item order is given in a standardized form. Since the order values are between 1 and I we use the standardized average 
of $\sum_{m=1}^M (o_{im}-1)/(M*(I-1))$ such that the standardized mean is in $[0,1]$. As for the misfit scores large values indicate that the item is not a Rasch item. The conclusions are the same as for the misfit scores, items 7,8,12 are problem items.

\begin{table}[h!]
 \caption{Misfit scores values for FIMS data} \label{tab:fimsind}
\centering
\small
\begin{tabularsmall}{llrrrrrrrcccccccccc}
  \toprule
   item&1 &2 &3 &4 &5 &5  &7 &8  &9  &10  &11 &12  &13 &14  \\ 
   misfit    & 0   & 0    &0    &0 &0.05  &0.20 &\textbf{0.85}  &\textbf{0.90} &0.25  &0.35     &0     &\textbf{1.00}  &0.35  &0.05\\
mean std &0.38 &0.19 &0.31 &0.38 &0.32 &0.51 &\textbf{0.86} &\textbf{0.87} &0.66  &0.48  &0.15  &\textbf{0.99}  &0.57  &0.34\\

\bottomrule
\end{tabularsmall}
\end{table}

\subsubsection*{Fears data}

We consider  data from the German Longitudinal Election Study (GLES), which is a long-term study of the German electoral process \citep{GLES}.  The data we are using  originate from the pre-election survey for the German federal  election  in  2017 and are  with political fears. The participants were asked: ``How afraid are you due to the ...'' - (1) refugee crisis?
 (2) global climate change?
 (3) international terrorism?
 (4) globalization?
 (5) use of nuclear energy?
 The answers were originally measured on Likert scales from 1 (not afraid at all) to 7 (very afraid) but have been transformed to binary responses.  It is unclear if  fear can be considered  the dominating unidimensional latent trait for these items. 
 
A tool that is often used to detect items that do not fit well is Andersen`s likelihood ratio test \citep{andersen1973goodness}. It  uses that item parameters can be  consistently estimated in any sample of persons provided the Rasch model holds in the corresponding population. It compares the estimates obtained in defined subgroups of subjects, which should only vary due to random fluctuations. If the test indicates that the model does not fit well Wald tests can be used to investigate which items are responsible and the result can be graphically illustrated, see, for example, \citep{mair2018modern}. Subgroups can be formed by external variables or on the basis of scale levels.    

\begin{figure}[h!]
\centering
\includegraphics[width=6cm]{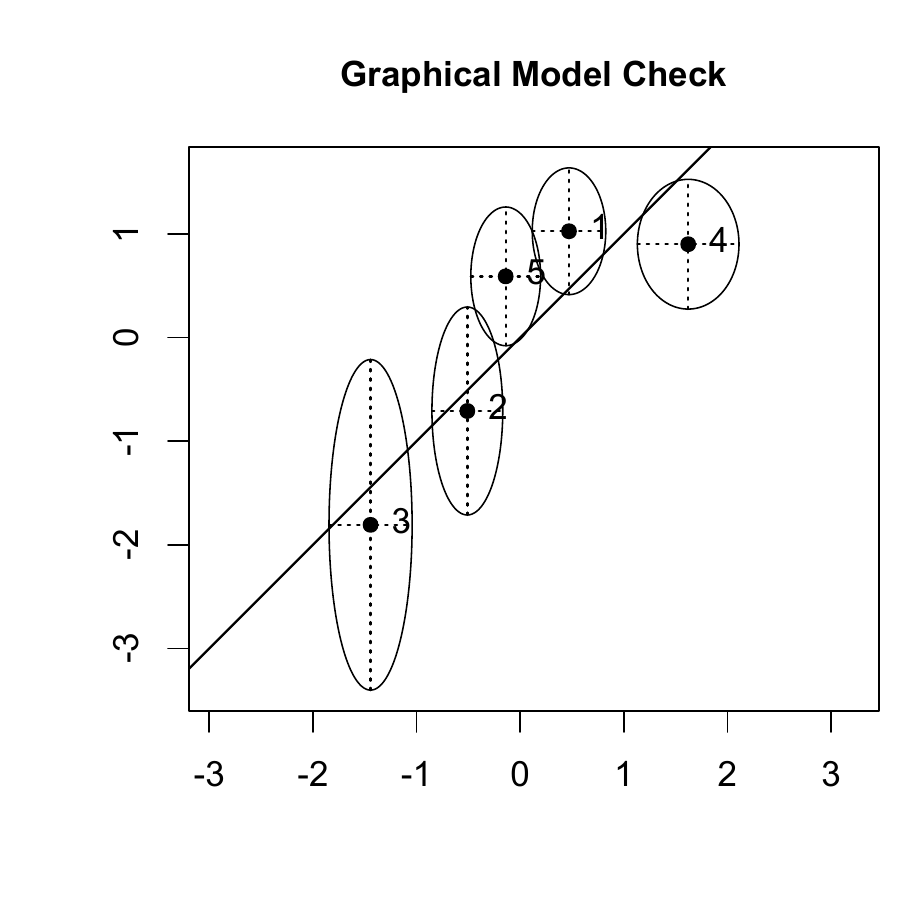}
\includegraphics[width=6cm]{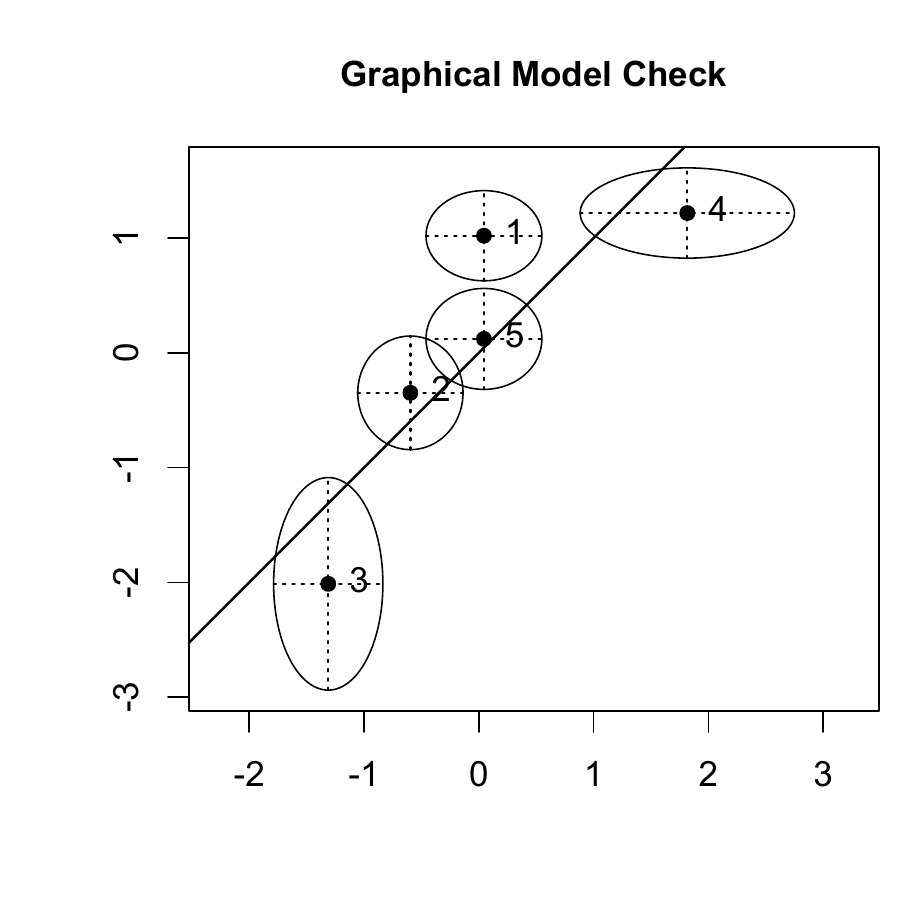}
\includegraphics[width=6cm]{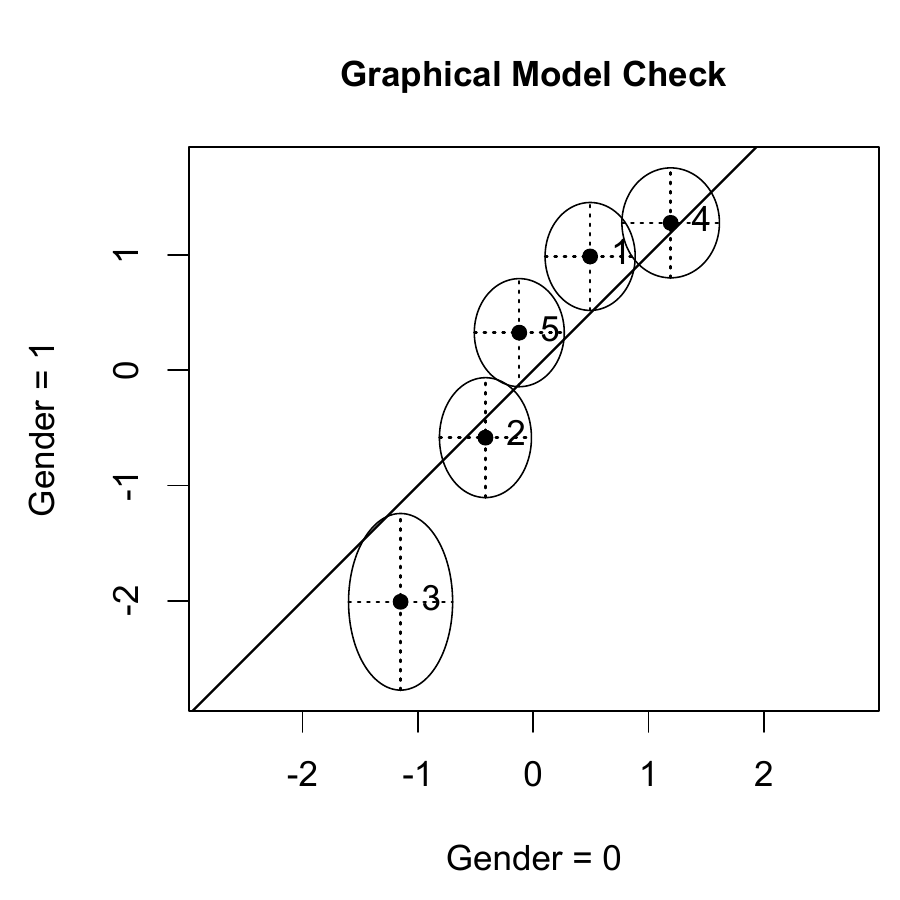}
\includegraphics[width=6cm]{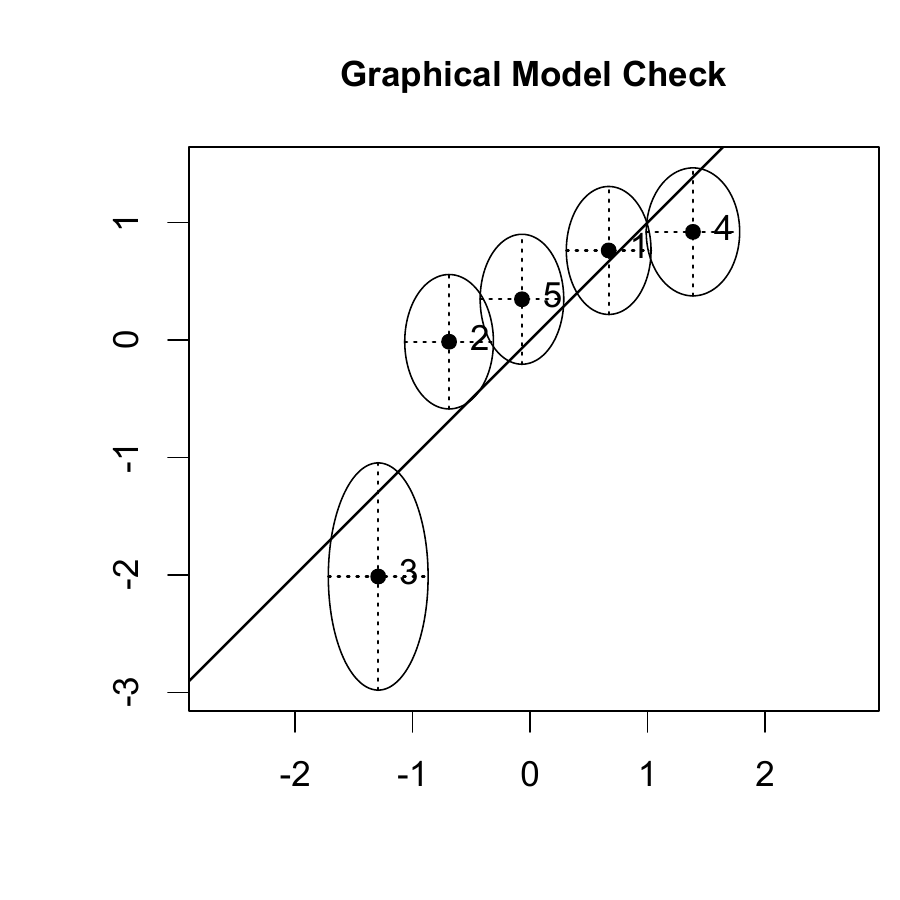}
\caption{Graphical model check for fears data, first row: subgroup split by  median and mean, second row: subgroup split by  gender and age }
\label{fig:fears-2}
\end{figure}

One problem is that the choice of the subsets crucially determines the diagnostic results.  Figure \ref{fig:fears-2} shows the results if subgroups are defined by the median of item scores, the mean of item scores, gender and age (dichotomized at 60). It is seen that it depends on the split criterion which  items  arouse suspicion. Even when comparing splitting by the median or the mean yields different results. When using the median for example item 5 looks suspicious but it is very close to the angle bisector when using the mean indicating good fit. Item 3 shows bad fit when using gender but fits rather well for the other split criteria.  
Although useful when investigating specific dependencies, for example on gender, the method is less convincing as a general item selection tool since it depends on the chosen subgroups and by using many alternate subgroups one can almost always find items that violate the assumption of the Rasch model.

\begin{figure}[h!]
\centering
\includegraphics[width=7cm]{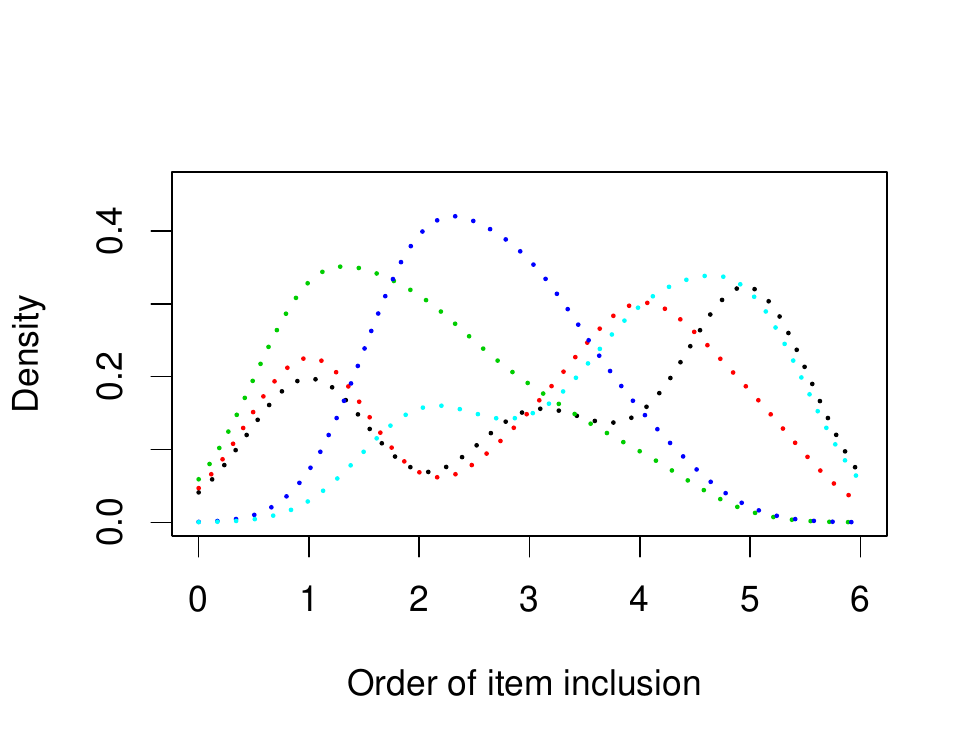}
\includegraphics[width=7cm]{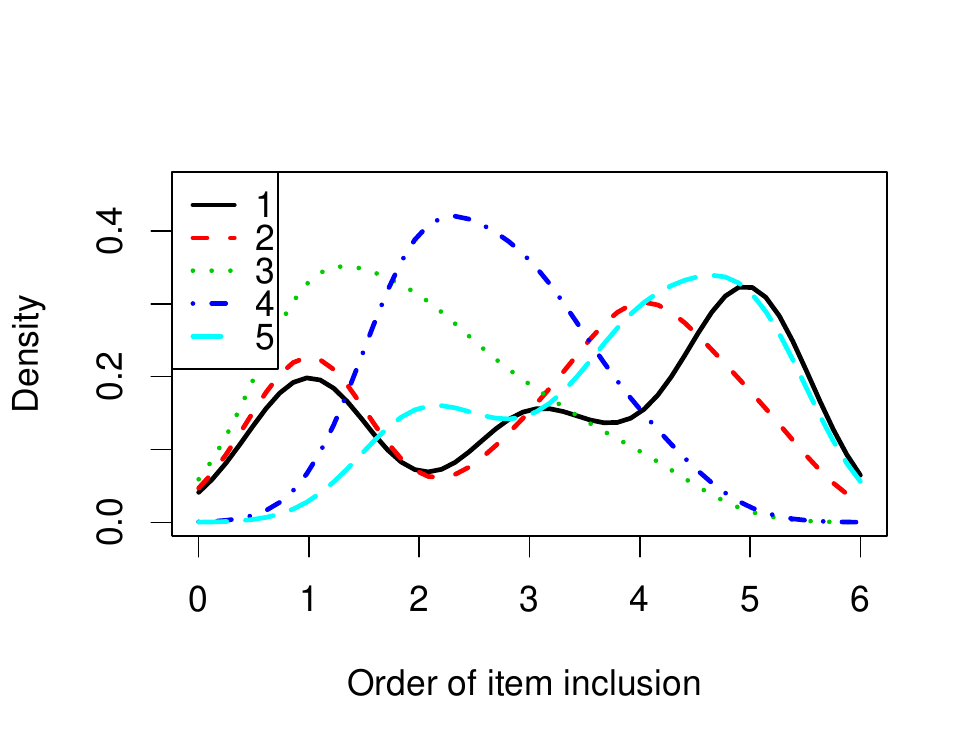}
\caption{Distribution of steps in which items are included for selected items (fears data)}
\label{fig:fearssel}
\end{figure}

An advantage of the clustering method is that it does not use the subgroup approach, which, in addition, only works for the Rasch model but not for alternative models to which the proposed approach can be extended. The sequential clustering procedure included the items in the order  3, 4, 2, 5, 1. Figure \ref{fig:fearssel} shows the order of item inclusion. 
Table \ref{tab:fearsind} shows the proportion of items that are included later than $0.75 \times I$ for all items. 
These values should be taken more seriously than the sequence of items itself since they are built as an average over  more data sets. The highest misfit is  
found for item 5 but as is seen from Figure \ref{fig:fearssel} this item has also been included in early steps in several data sets. Thus, there is no strong evidence that it is a non-Rasch item although it can be considered as the 'weakest' item in the set of items.

\begin{table}[h!]
 \caption{Misfit scores values for fears data} \label{tab:fearsind}
\centering
\begin{tabularsmall}{llrrrrrrrcccccccccc}
  \toprule
   item&1 &2 &3 &4 &5 \\ 
   misfit score    &0.52 &0.54  &0.10 &0.14  &0.70\\
   mean std &0.77 &0.55 &0.41 &0.50 &0.75\\
\bottomrule
\end{tabularsmall}
\end{table}

\subsection{Item selection and alternative methods }
 
The basic idea to select items with a common trait that is used here is to exploit that the presence of non-Rasch items will affect estimation results. If one has a set of items, for which the Rasch model holds and one excludes  items or includes  new items for which the Rasch model also holds estimates should not change strongly. That holds for all parameters, the difficulty parameters as well as the standard deviation of the mixing distribution. In our selection procedure we used the standard deviation. The algorithm did not specify explicitly that changes are considered but it implicitly does. If a sequence of items $C_1^{(\ell)}$ from clustering $P_{\ell}$ is given one computes the  standard deviations of the potential 
clusters $C_1^{(\ell)}\cup C_j^{(\ell)}$, $j=1,\dots,I-\ell$, $s_{ij}=\sigma_{\theta_p}(C_1^{(\ell)}\cup C_j^{(\ell)})$ and selects the clustering that has the largest value. It is equivalent to computing the \textit{changes} of estimated standard deviations 
\[
\text{ch}_{ij}= \sigma_{\theta_p}(C_1^{(\ell)})-\sigma_{\theta_p}(C_1^{(\ell)}\cup C_j^{(\ell)}),
\]
where $\sigma_{\theta_p}(C_1^{(\ell)})$ is the estimated standard deviation if the model is fitted for items $C_1^{(\ell)}$, and choosing that clustering  that shows the \textit{minimal} change. It is equivalent to the original algorithm since the cluster $C_1^{(\ell)}$ is the same in all
clusterings.

The basic idea, namely to investigate the change in parameters when new items are included, can also use the change in difficulty parameters. 
Then for a given sequence of items $C_1^{(\ell)}$ from clustering $P_{\ell}$  one computes the  change 
\[
\text{ch}_{ij}= \sum_{i=1}^{c_1} \delta_i(C_1^{(\ell)})-\delta_i(C_1^{(\ell)}\cup C_j^{(\ell)}),
\]
where $\delta_i(M)$ denotes the estimated difficulty parameters when the model is fitted for items $M$, and $c_1$ is the number of parameters in $C_1^{(\ell)}$.
Then one select that clustering that shows the \textit{minimal} change. The method exploits that item parameter estimates should not change too much if Rasch compatible items are included in the item set. 

The method yields very similar clusters as the standard deviation method if one uses in the first selection step where pairs of items are examined the standard deviation criterion. Thus a hybrid version works well.
The pure algorithm that uses the change in difficulty parameters in all steps performs poorer since in the first step there is a tendency to include wrong items. The advantage of the standard deviation method is that it is based on a single criterion. A possible advantage of the difficulty parameter approach in its hybrid version is that it can also be used if the mixing distribution is not the normal distribution while modifications are necessary if one wants to use the standard deviation approach.

\section{Hierarchical clustering }

In the following we consider the more challenging problem of finding clusters of items from a more diverse set of items. It is suspected that different traits are at play and subsets of items are determined by different traits. The objective is to cluster items such that   clusters are unidimensional.  

\subsection{Hierarchical clustering based on marginal estimates}

The extended clustering method also uses   the 'homogeneity' measure $S(P,P_{ij})= \sigma_{\theta_p}(C_i \cup C_j)$ given in equ. 
(\ref{equ:hetf}), which should take large values in the range of the true standard deviation if items in the clusters to be fused $C_i \cup C_j$ are Rasch items.
However, in contrast to the sequential inclusion of items to form one big cluster
now any two item sets can be fused to build a new cluster instead of enlarging one specific cluster. The result of the algorithm is  hierarchical clustering.

\vspace{0.5cm}
\hrule
\begin{center}{\textbf{Clustering II: Hierarchical Clustering}}\end{center}

\begin{description}
\item{\textit{Step 1 (Initialization)}}
 
 As in previous algorithm.

\item{\textit {Step 2 (Iteration)}}

For $\ell=1,\dots,I-2$

\begin{itemize}
\item[(a)] Computation  

Let  $P_{\ell ij}$ denote the partition that contains the  clusters $C_i^{(\ell)}\cup C_j^{(\ell)}$, $i=1,,\dots,I-\ell$, $j=2,\dots,I-\ell$,
which is built from  $P_{\ell}$ by a fusion of $C_i^{(\ell)}$ and $C_j^{(\ell)}$. 
Compute the  corresponding estimated standard deviations   $s_{\ell ij}=S(P_{\ell},P_{\ell ij})$

\item[(b)] Selection

Select the partition with the largest standard deviation $s_{\ell ij}$,
\[
(i_0,j_0) = \operatorname{argmax}_{(i,j)} s_{\ell ij},
\]

\item[(b)] Cluster fusion

Form the new partition $P_{\ell+1}$ by fusing the clusters $C_{i_0}^{(\ell)}$, $C_{j_0}^{(\ell)}$.
\end{itemize}

\hrule 
\vspace{0.5cm}
\end{description}

Hierarchical clustering is a topic that has extensively covered in the cluster literature, see, for example, \citet{romesburg2004cluster}, \citet{hennig2015handbook}. Traditional  cluster analysis uses distances computed from vector-valued observations, and clusters are determined by looking for clusters such that distances between observations within one cluster are small but distances between clusters are large. These methods can be used in a straightforward way by defining distances between  items by $d_{ij}= d(Y_{i.}, Y_{j.})$, where  $Y_{i.}^T=(Y_{i1},\dots,Y_{iP})$ is the vector of observed responses for item $i$ and $d(.)$ is a distance measure as the Euclidean distance.

The method considered here is quite different.  Instead of using distance measures it  fuses those clusters for which the standard deviations of clusters that are considered for fusion are large.
In Section \ref{sec:id} we used the hit rate as a measure of performance. That was possible because it was assumed that most of the items do share a common trait, the other items were considered as non-Rasch items. In the present case there is no distinction between Rasch items and non-Rasch items. Instead it is assumed that there are clusters of items such that the items of each cluster follow a Rasch model but different traits are measured by the clusters. Thus different performance measures are needed.   

The performance of a hierarchical clustering procedure can be evaluated by computing if pairs of item that are in one cluster in the true partition are also found in one cluster in the estimated partition with a fixed number of clusters. More concrete, we consider  ``hits'' and ``false'' allocations defined  by

{\scriptsize
\[
h_j = \frac{\text{ Number of pairs of items that are in one cluster in the true partition as well as in the estimated partition with j clusters}}
{\text{Number of pairs of items that are in one cluster in the true partition}},
\]
\[
f_j = \frac{\text{ Number of pairs of items that are not in one cluster in the true partition but in the estimated partition with j clusters}}
{\text{Number of pairs of items that are not in one cluster in the true partition}}.
\]
}

In the ideal case the hit rate would be 1 and the number of false allocations 0. Of course this could only occur in the case where the number of clusters coincides with the number of true classes. In all other cases these ideal numbers can not be reached. To obtain a wider picture of the performance it is advisable to examine the misfit scores for all clusterings, that means  
hit rates and false rates are shown together  yielding a visualization similar to ROC curves as considered, for example, by \citet{Pepe:2003}. 
The sequence of hit and false rates $(h_1,f_1),\dots,(h_I,f_I)$ yields a curve with end points $(h_1,f_1)=(1,1)$ and $(h_I,f_I)=(0,0)$.
The closer the curve is to the left upper corner the better the performance of the clustering procedure.

\subsection{Illustration }
Let the true partition be given by $C_1=\{1,\dots,6\}$ and $C_2=\{7,\dots,12\}$. Thus different Rasch models hold  for items 1 to 6 and items 7 to 12 (item parameters as in Section \ref{sec:marg}). For both models person parameters are drawn from a standard normal distribution ($P$=200).

\begin{figure}[h!]
\centering
\includegraphics[width=7cm]{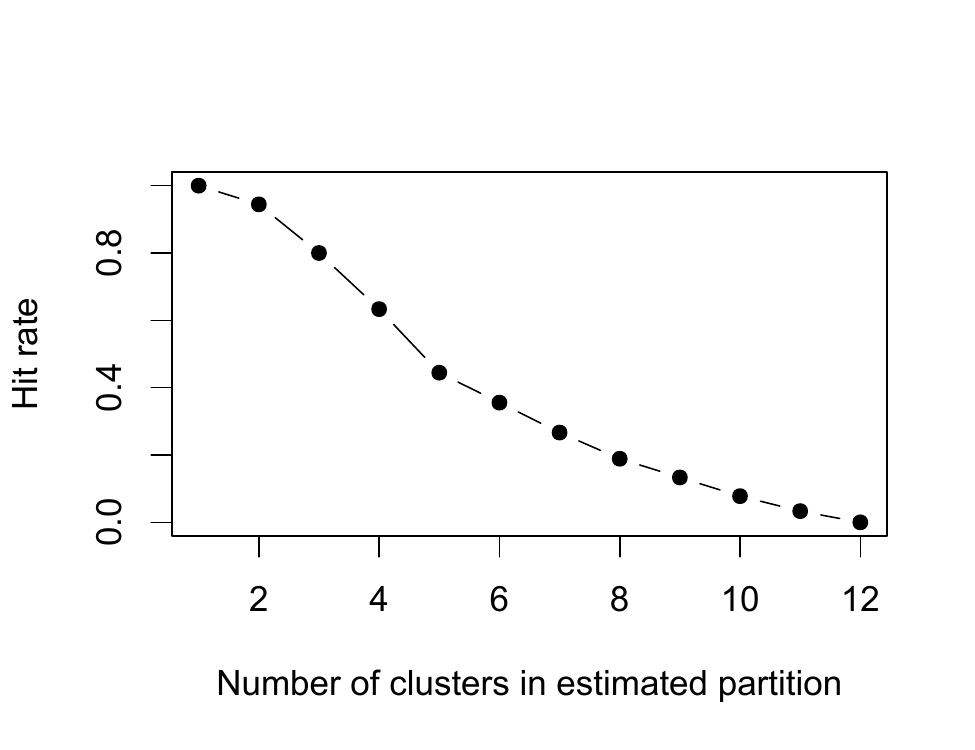}
\includegraphics[width=7cm]{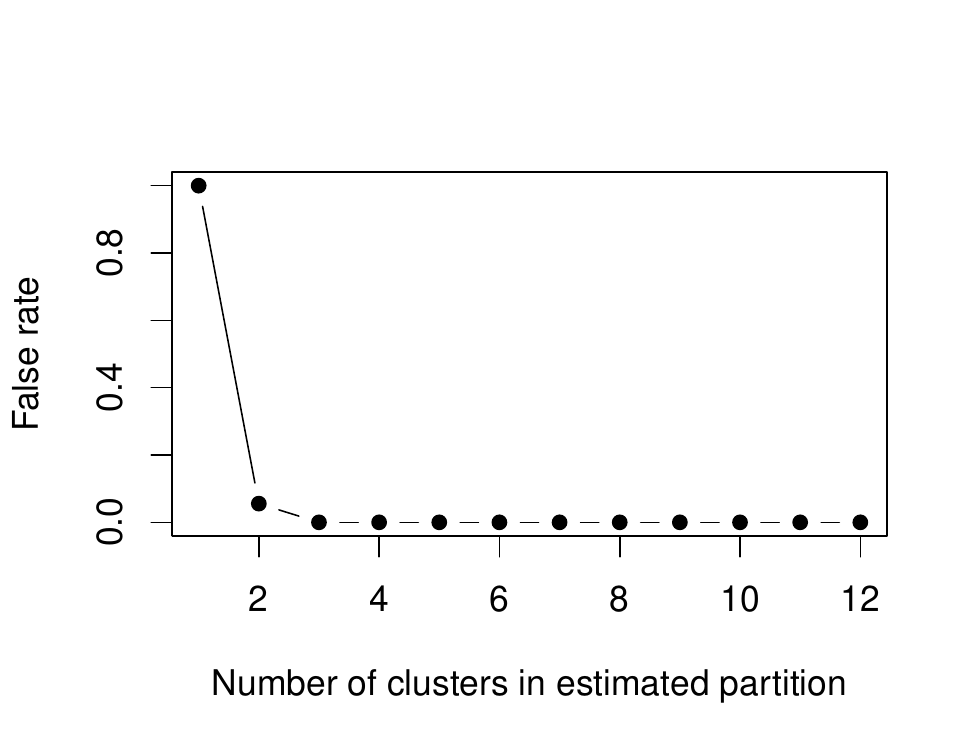}
\includegraphics[width=7cm]{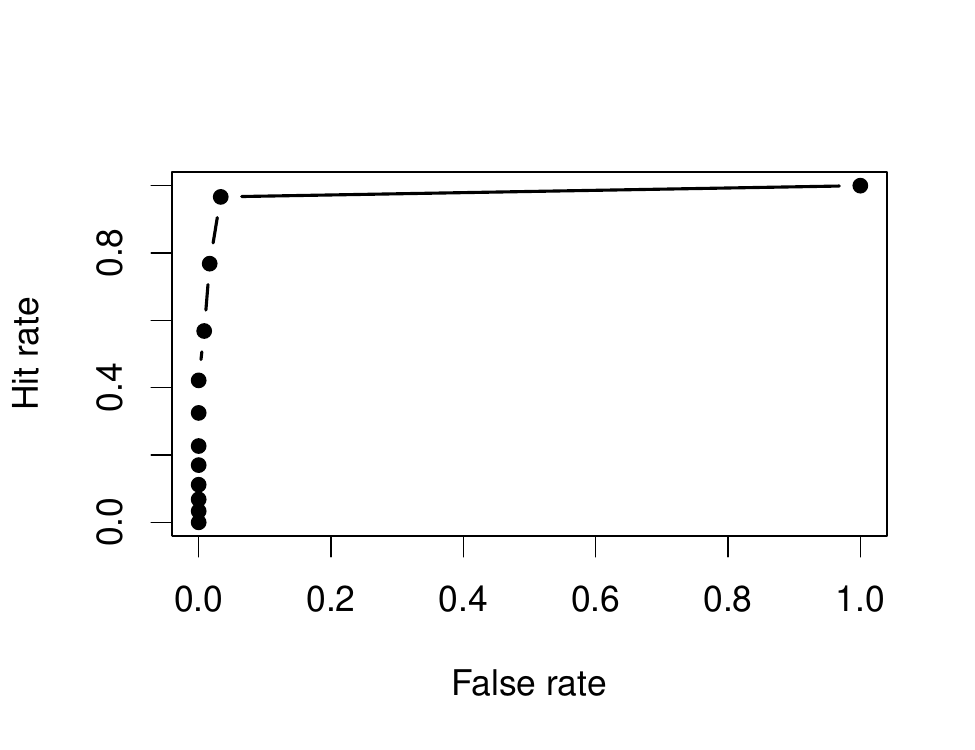}
\includegraphics[width=7cm]{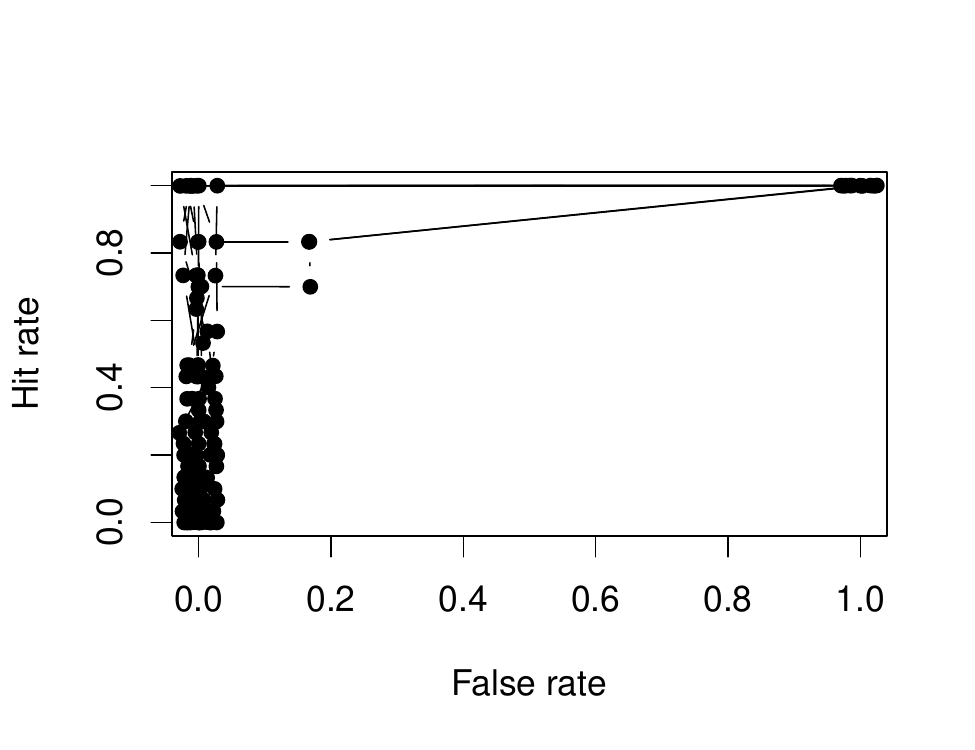}
\caption{Hit   and false allocation rates for 12 items with true clustering $\{1,\dots,6\},\{7,\dots,12\}$}
\label{fig:hiertwelve1}
\end{figure}

Figure \ref{fig:hiertwelve1} shows the resulting hits and false allocations. It is seen that  for small numbers of clusters, which corresponds to the true clustering,  the number of hits is  in the estimated partition is rather large and the number of false allocations is small, which indicates that the clustering procedure performs rather well. The 'ROC curve' (third picture in Figure \ref{fig:hiertwelve1}) is very close to point  (1,1) for two classes. For larger numbers of classes the hit rate decreases slowly but the false allocation rate remains very small. In the first steps of the algorithm items that come from the same model are also found in the same cluster and items that are from different models are not found in the same cluster. The last picture shows hit rates plotted against false rates for the first ten simulations (slightly jittered). It is seen that in many simulations the two cluster solution means hit rate 1 and false allocation rate 0.

\subsection{Comparison with traditional cluster procedures }

\begin{figure}[h!]
\centering
\includegraphics[width=7cm]{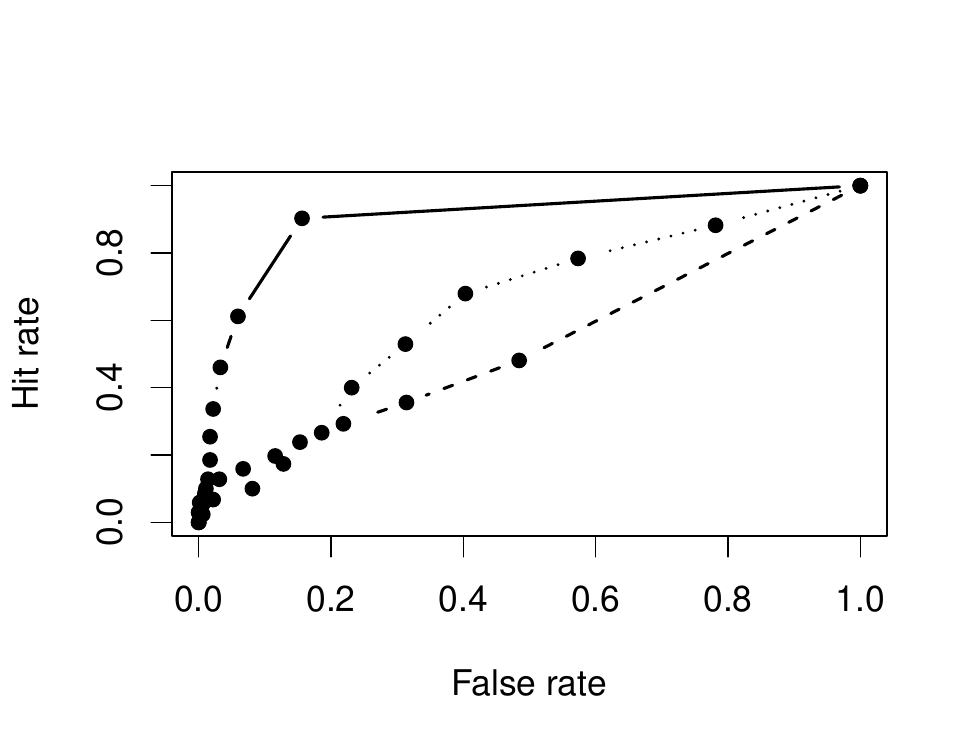}
\includegraphics[width=7cm]{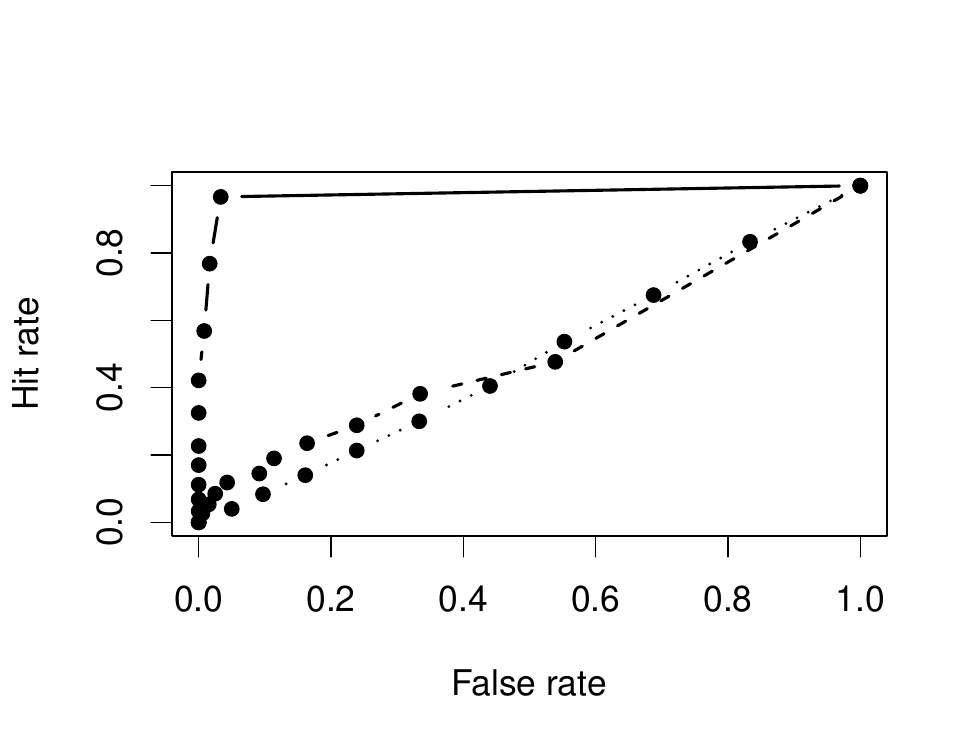}
\caption{Comparison classical clustering (dashed: average linkage, dotted: centroid clustering) for 12 items, true clusters $\{1,\dots,8\}, \{9,\dots,12\}$ (left) and $\{1,\dots,6\}, \{7,\dots,12\}$ (right);  solid line is proposed cluster procedure ($P=200$)}
\label{fig:comp12-1}
\end{figure}

As mentioned previously traditional cluster methods can be used on the distances between responses on items. In the following the cluster method based on marginal estimates is compared to traditional methods. As reference we consider the average distance method and the centroid method provided by the R function \textit{hclust}.  Figure \ref{fig:comp12-1} shows the performance in terms of hits and false allocations over all built clusters for the case of 12 items with the true clusters   given by items $\{1,\dots,8\}$ and $\{9,\dots,12\}$ (solid line: proposed cluster procedure, dashed: average linkage, dotted: centroid clustering ).
It is seen that the cluster procedure based on marginal estimates distinctly outperforms the traditional methods.

\begin{figure}[h!]
\centering
\includegraphics[width=7cm]{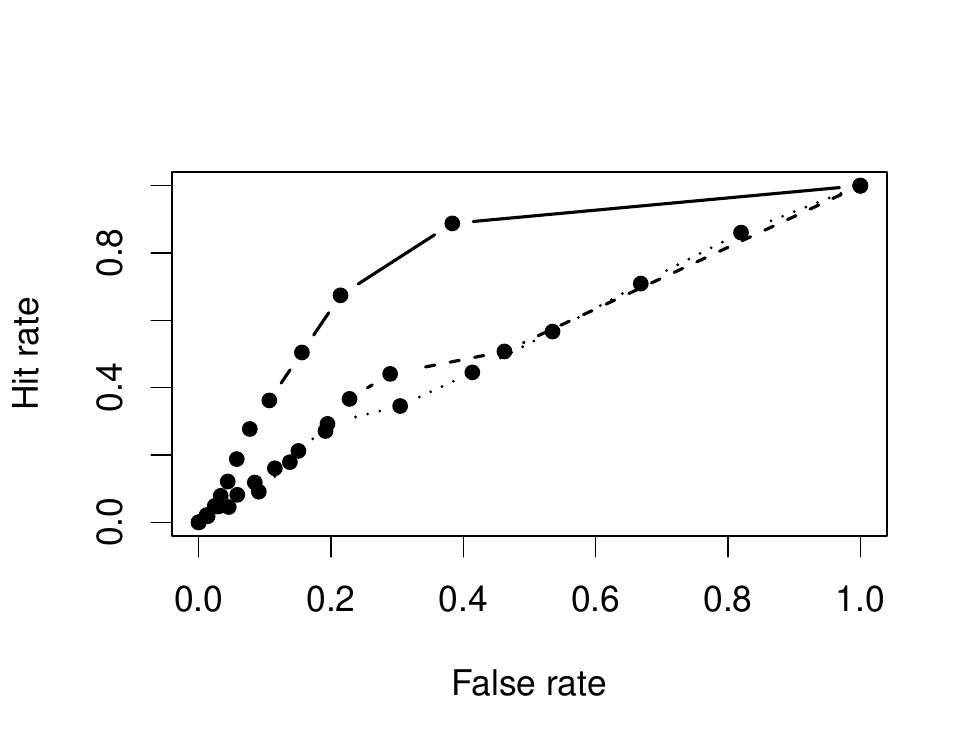}
\includegraphics[width=7cm]{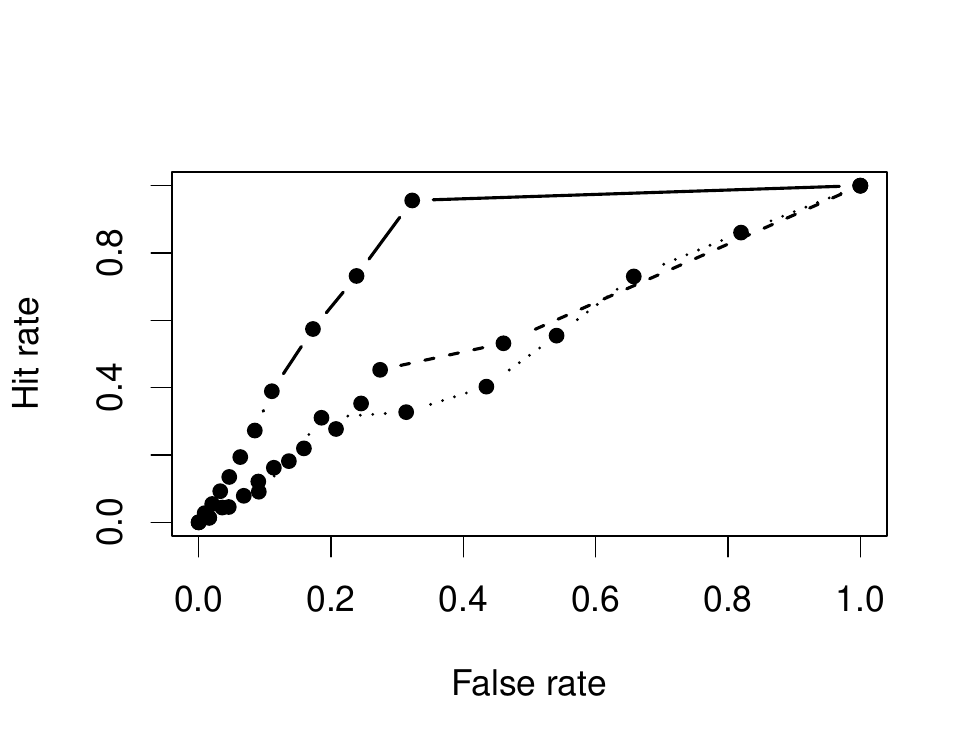}
\caption{Comparison classical clustering (dashed: average linkage, dotted: centroid clustering) for 12 items, true clusters $\{1,2,3,4\},\{5,\dots,10\}\{11,12\}$ (first) ;  solid line is proposed cluster procedure, left: $P=200$, right: $P=400$}
\label{fig:comp6morecl}
\end{figure}

Figure \ref{fig:comp6morecl} shows the performance of classical methods and the new method for 12 items with three true clusters given by $\{1,2,3,4\},\{5,\dots,10\}\{11,12\}$. It is seen that the marginal estimate method performs definitely better than traditional methods.

\subsection{Homogeneity for single item sets }

Although the identification of homogeneous item sets works well on average in applications the procedure yields just one hierarchical clustering, which can be seen as exploratory but may  not be very reliable as an indicator of the true item structure. As in the case of sequential clustering we use a stability criterion that is based on an average of cluster buildings.

For one data set repeatedly subsets of persons  are drawn ( used  proportion is 0.5). For each subset the hierarchical cluster procedure is run. Then, it is computed how often pairs of items are found in the same cluster over all clustering steps. This is averaged over the drawn subsets. More concrete, we consider for the each subset of persons
\[
s_{ij}= \sum_{\ell=1}^{I-1}  \sum_{r=1}^{I-\ell} I((i,j)\in C_r^{(\ell)}) /(I-1), 
\]
where $I(.)$ is the indicator function. The sum is over the clusterings $C_r^{(\ell)}$, $\ell=1,\dots, I-1$, which means the final clustering, which consists of just one set containing all items, is omitted. $s_{ij}$  measures how often the pair $(i,j)$ is found within the same cluster over the whole clustering (divided by the number of cluster levels). 
Averaging over the $M$ drawn subsets yields the final similarity measures  for pairs of items.
If the similarity for a pair of items is large it indicates that the items share the same trait.   

Although pairs of items reflect only a part of the complexity of the various clusterings they are considered since they are easy to access.
More complex measures of the stability of clusters as an average over various cluster could certainly be constructed but at the cost of easiness of interpretation. 

As an example we consider the stability of clusters for 6 items with the true clustering $\{1,2,3\},\{4,5,6\}$. The average over 50 simulated data sets (15 randomly drawn subsets for each data set) yielded the similarities given in Table \ref{tab:simstab}. 
It is seen that for pairs of items from $\{1,2,3\}$ similarities are large. The same holds for pairs of items from $\{4,5,6\}$. In contrast similarities are small if pairs are built with one item from $\{1,2,3\}$ and the other from $\{4,5,6\}$. 

\begin{table}[h!]
 \caption{Similarities for 6 items with $\{1,2,3\},\{4,5,6\}$} \label{tab:simstab}
\centering 
\begin{tabularsmall}{llrrrrrrrcccccccccc}
  \toprule
   item&1 &2 &3 & &4 &5 &6\\ 
   1     &1.00 &0.34 &0.36 &&0.06 &0.07 &0.07\\
   2     &0.34 &1.00 &0.36 &&0.09 &0.12 &0.11    \\
   3     &0.36 &0.36 &1.00 &&0.09 &0.15 &0.08                     \\
   \\
   4     &0.06 &0.09 &0.09 &&1.00 &0.36 &0.32                \\                    
   5     &0.07 &0.12 &0.15 &&0.36 &1.00 &0.32      \\
   6     &0.07 &0.11 &0.08 &&0.32 &0.32 &1.00      \\

\bottomrule
\end{tabularsmall}
\end{table}

\begin{figure}[h!]
\centering
\includegraphics[width=7cm]{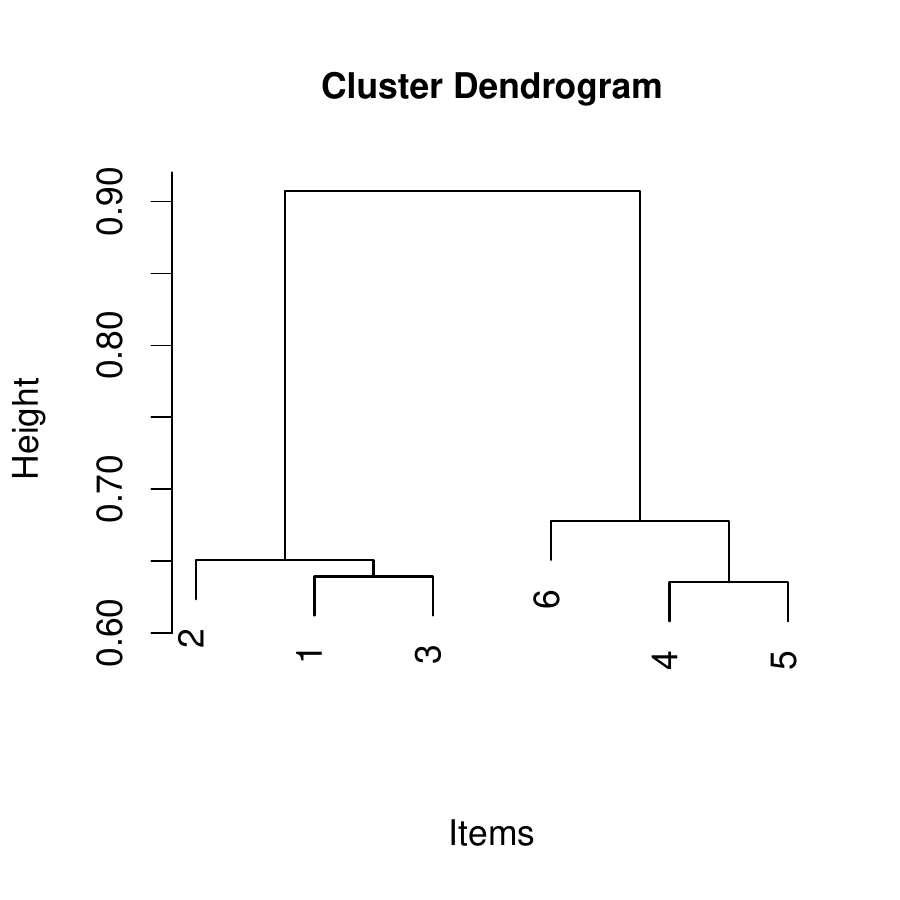}
\caption{Dendrogram for pairwise distances if true clusters were $\{1,2,3\},\{4,5,6\}$ }
\label{fig:pairav}
\end{figure}

For exploratory purposes it is helpful to transform the similarities into distances by defining $d_{ij}=1-s_{ij}$. Then one can use standard cluster methods to investigate the cluster structure based on these averaged distance measure. Figure \ref{fig:pairav} shows the resulting dendrogram if average linkage is used. It is seen that the true dendrogram reflects the true cluster structure very well. 

\section{Further illustrating examples }

\subsection{Fears data }

\begin{figure}[H]
\centering
\includegraphics[width=7cm]{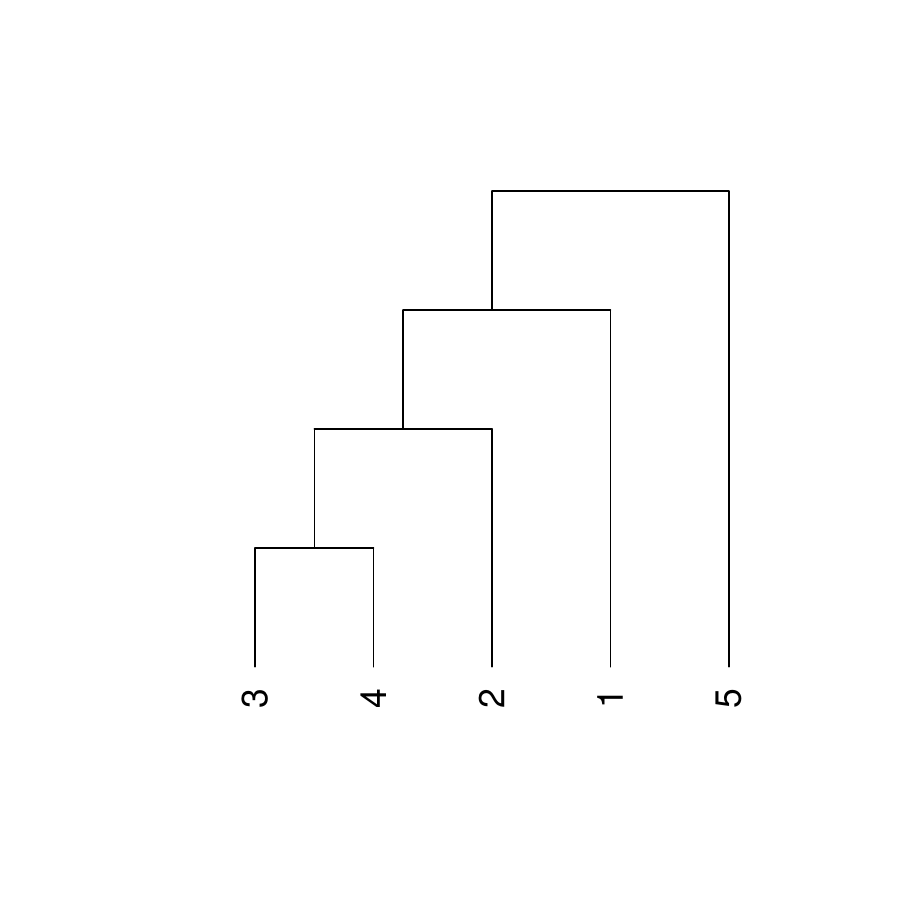}
\caption{Dendrogram for fears data }
\label{fig:fearsav}
\end{figure}

We consider again the fears data  from the German longitudinal election study from Section \ref{sec:ill} but now use the hierarchical cluster procedure.
The resulting clusters are shown in  the dendrogram given in Figure \ref{fig:fearsav}. It is seen that one cluster is grown by successively including items although the hierarchical procedure was used.  
Interestingly, the clusters that are built are almost the same as in the sequential clustering procedure used before.
The sequential procedure  yielded the item sequence 3, 4, 2, 5, 1 while the hierarchical yielded 3, 4, 2, 1, 5. But as has already been argued in the sequential clustering procedure it is more important to look at the misfit scores, which show that, although item 5 is  included   later than the other items, there is no strong evidence that it is a non-Rasch item. It seems that most of the items can be considered as being determined by one trait.

\subsection{Discalculia }

\begin{figure}[h!]
\centering
\includegraphics[width=7cm]{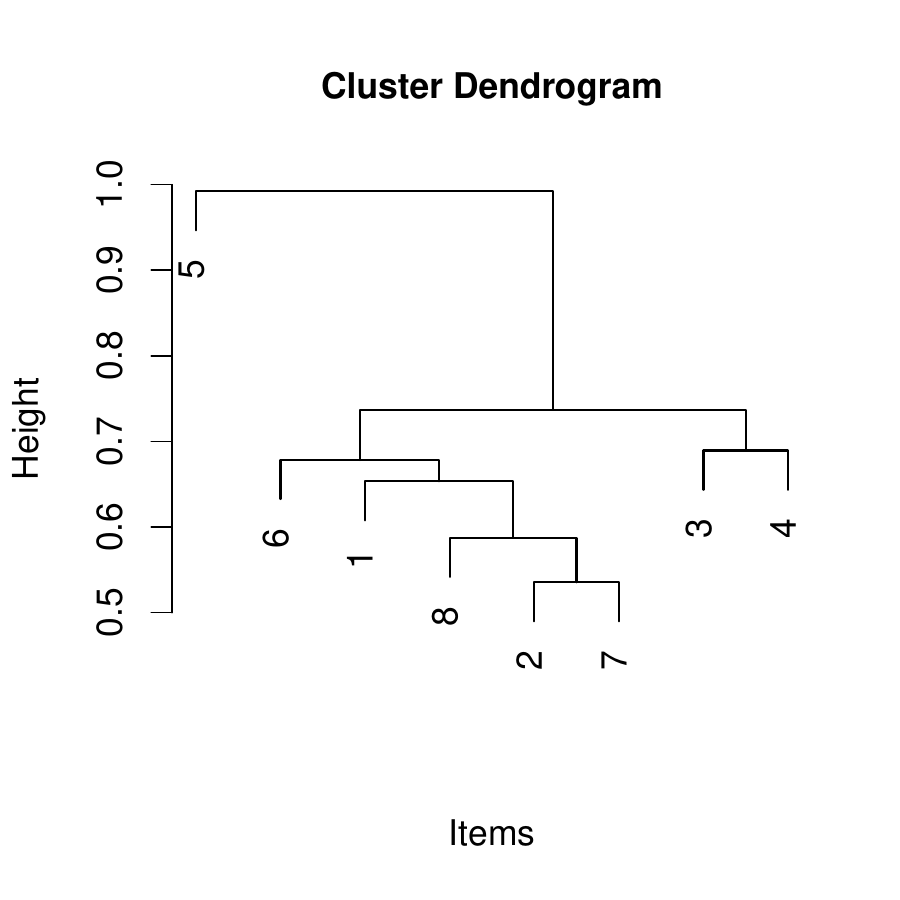}
\caption{Dendrogram for pairwise distances of discalculia data}
\label{fig:pairavzar}
\end{figure}

We use the data set \textit{ZAREKI-R} from the R package \textit{MPsychoR} \citep{mair2018modern}. It is based on a Neuropsychological Test Battery for Number Processing and Calculation in Children \citep{vonAA} for the assessment of dyscalculia in children. The number of children in the sample is $P =341$.  As \citet{mair2018modern} we consider the eight binary subtraction items.
Hierarchical clustering yields similarities and corresponding distances that are used to draw the dendrogram (average distance measure) given in 
Figure \ref{fig:pairavzar}. It is seen that item 5 is distinctly separated and can hardly be expected to have the same common trait as the other items. This is in accordance with the findings of \citet{mair2018modern}, who also concludes that item 5 is  differs from the other items by using quite different investigation methods.

The dendrogram suggests that there might  be two further  clusters that can be separated formed by items 3,4 and 1,2,6,7,8.
For further analysis  we  used the sequential cluster procedure to investigate the homogeneity of items. The misfit indicator values if all items are considered as given in Table \ref{tab:zar} again shows that item 5 is standing out having quite large misfit values. It should certainly be excluded. The misfit values in the reduced item set without item 5 still seems not very homogeneous since in particular the values for item 4 are still very large (second part of Table \ref{tab:zar}). The dendrogram suggests that 1,2,6,7,8 form a rather homogeneous set. The corresponding misfit values (third part of Table \ref{tab:zar}) underpin this proposition. There are no items with extreme misfit values.

\begin{table}[H]
 \caption{Indicator values for discalculia data } \label{tab:zar}
\centering
\begin{tabularsmall}{llllllllcccccccccc}
  \toprule
   item&1 &2 &3 &4 &5 &6 &7  &8\\ 
\midrule    
misfit &0    &0 &0.07 &0.43 &0.97 &0.43    &0  &0.1                 \\
mean &0.33  &0.3 &0.47 &0.68 &0.95 &0.65  &0.2 &0.43\\

\midrule
misfit &0.08 &0.08 &0.28  &0.8 &- &0.60    &0  &0.18                 \\
mean std &0.4 &0.36 &0.55 &0.83    &- &0.72 &0.16 &0.48\\

\midrule
misfit &0.63 &0.20 &- &- &- &0.50 &0.20 &0.47                 \\
mean std &0.67 &0.33 &- &- &- &0.57 &0.38 &0.55\\

\bottomrule
\end{tabularsmall}
\end{table}

\section{Concluding remarks }
The proposed item selection procedure has been shown to work well for Rasch models. However, it can be extended in a straightforward way to item selection in more general models as the 2PL model or ordinal models as the  partial credit model \citep{Masters:82, MasWri:84}, the graded response model \citep{samejima1995acceleration,samejima2016graded} and the sequential model \citep{Tutz:90b}, since for all of these models  marginal estimation can be used.

The sequential selection method can also be used with an anchor item, which is known  to measure the latent trait of interest. 
Then a set of items is selected that shares the same latent trait. The method with an anchor item also applies if the set of items is less structured, that is, one suspects that items with quite different traits are in the set of available items.

The hierarchical method, although showing good performance on average, is considered an exploratory tool. As any hierarchical cluster method it offers a view on the similarities of objects but gives no final answer to the question how many clusters and which ones are really in the data. It is worth mentioning that also cluster methods that rely on test statistics are not able to give final answers. Typically many tests have to be performed and the overall error rate is not controlled. One is far from the situation of an experimentum crucis where the error rates are known and the use of test statistic might yield a false sense of security. 
Nevertheless, the dendrograms give some information on  which items can be expected to share a common trait. Within the given framework items that seem to form clusters can be investigated  by using the sequential cluster method  to investigate if they actually form a homogeneous item set.   

In principle, the sequential cluster method aims at selecting items that share a latent trait, the  structure of the other items is not investigated. They may be generated by different traits with  possibly different response functions or by multidimensional models. Realistically the true data generating process for items can rarely be expected to  strictly follow a unidimensional item response model. Therefore, items will be selected that are  (approximately) unidimensional but also items might be included that are (approximately) multidimensional but  have a strong trait component that is the same one as in the selected 'unidimensional ' items. 
In the  hierarchical clustering procedure it is explicitly assumed that different traits are present and it is searched for clusters of items that are 
approximately unidimensional although other dimensions might be involved to a lesser degree. An approach to hierarchical clustering that explicitly uses  multidimensional models but also assumes  that the response is affected by only one of the latent variables has been proposed by \citet{bartolucci2007class}. However, it is a latent class modeling approach, in which the latent traits are represented through a random vector with a discrete distribution,  and the number of classes has to be kept rather low for the model to be identifiable.

\bibliography{literatur}

\section{Appendix }

\subsection{Identifying non-Rasch items }

\begin{figure}[H]
\centering
\includegraphics[width=7cm]{hit12-3}
\includegraphics[width=7cm]{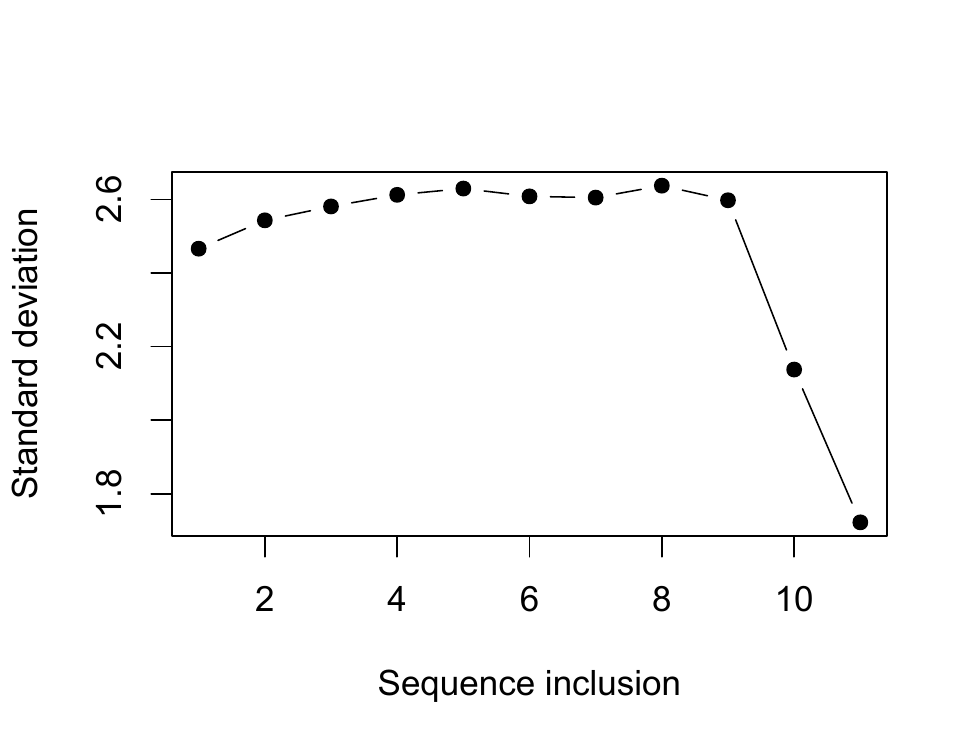}
\caption{Hit rate, standard deviation for 12 items, items 11 and 12 are non-Rasch itemsrep=1, $P=200$, $\sigma_{\theta}=3$}
\label{fig:hettw3}
\end{figure}

In the following we show further results for clustering strategy I, which aims at identifying non-Rasch items in a given set of items.
Figure \ref{fig:hettw3} shows the hit rates and the estimated standard deviations for 12 items where items 11 ann 12 are non-Rasch items if the standard deviation of the mixing distribution is larger than in previous simulations. It is seen that selection is perfect and the decrease in estimated standard deviations is very pronounced when non-Rasch items are included. 
Figure \ref{fig:hettw3} shows the results for the  case where the amount of information in the data is much smaller, the  
number of persons is just 100 and the true standard deviation is 0.6.  Even in this case the hit rates are close to 1, but the decrease in estimated standard deviation is less distinct.

\begin{figure}[H]
\centering
\includegraphics[width=7cm]{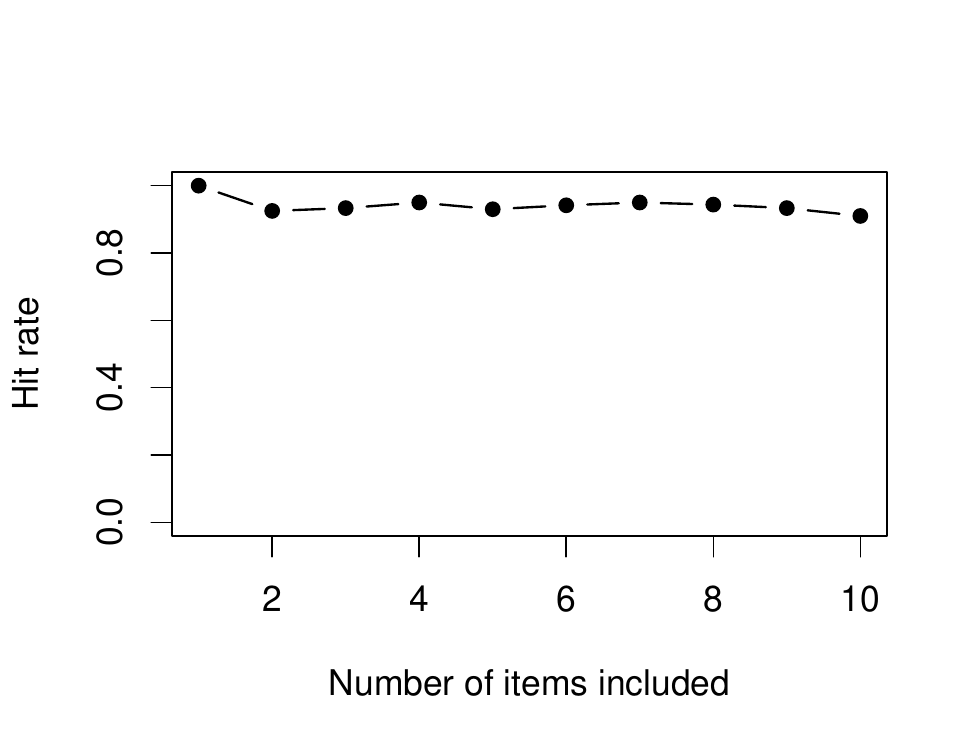}
\includegraphics[width=7cm]{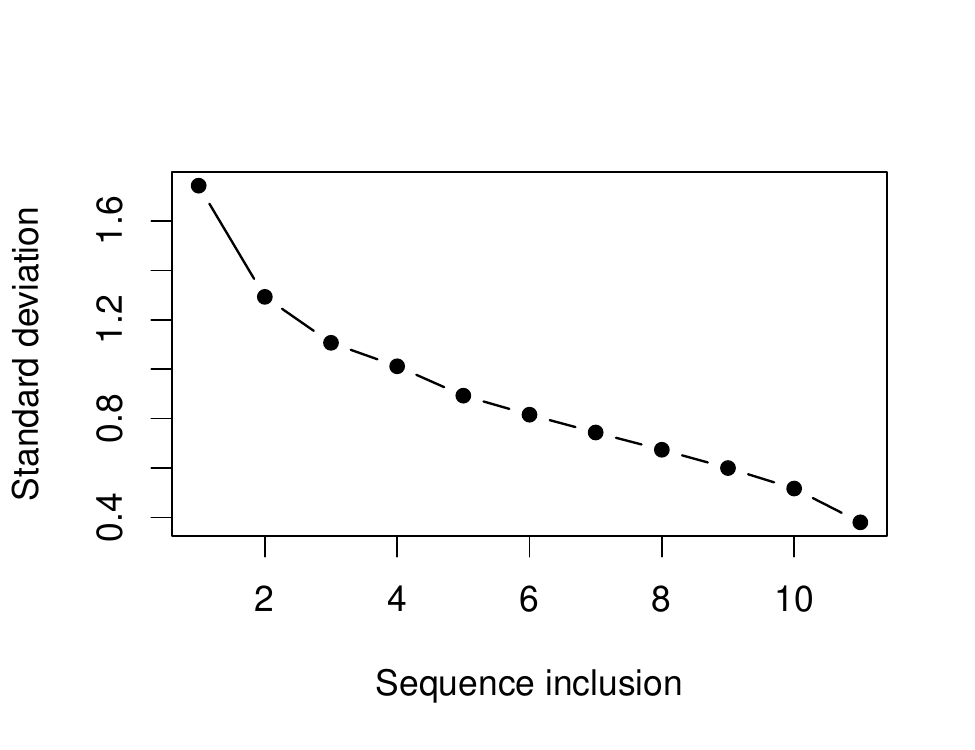}
\caption{Hit rate, standard deviation for 12 items, items 11 and 12 are non-Rasch items, $P=100$, $\sigma_{\theta}=0.6$}
\label{fig:hettw4}
\end{figure}

\begin{figure}[H]
\centering
\includegraphics[width=7cm]{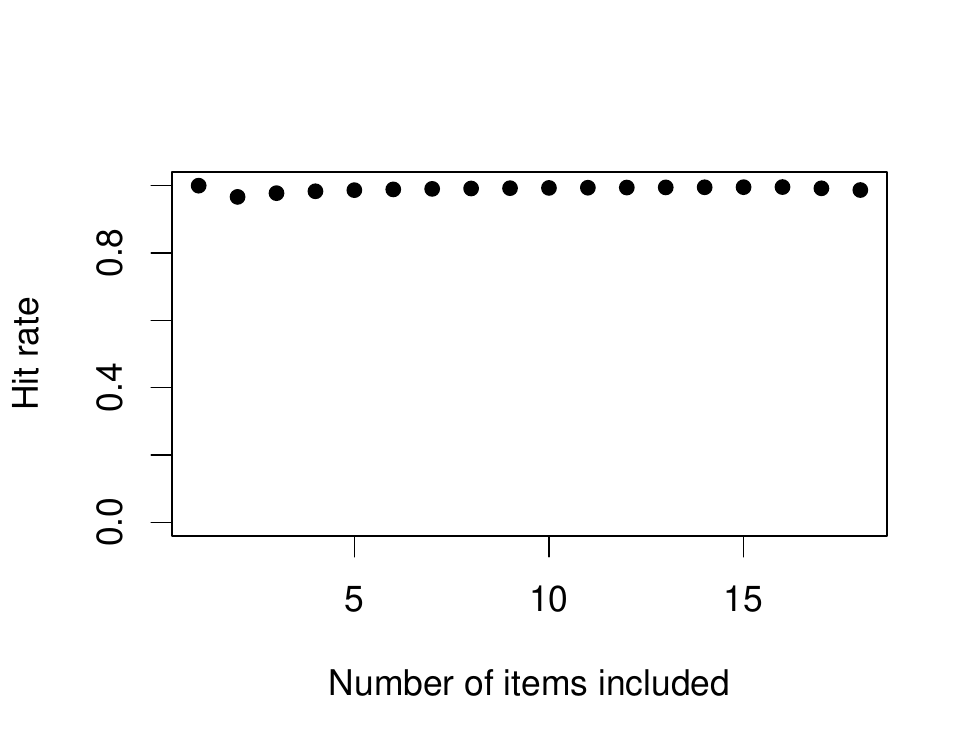}
\includegraphics[width=7cm]{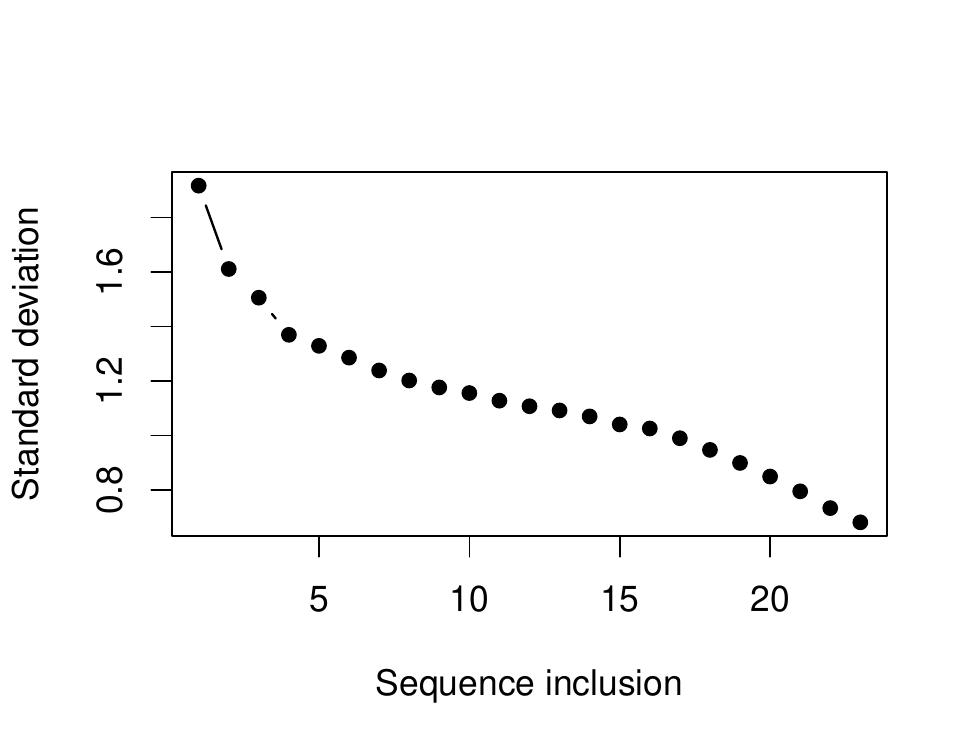}
\caption{Hit rate and heterogeneity for 24 items, 18 Rasch items and 6 non-Rasch items}
\label{fig:hier24}
\end{figure}

Figure \ref{fig:hier24} shows  results for a larger sets of items.
The number of items was 24, 18 items were Rasch items, the 
 first six items have difficulty parameters 0, -1.5, -1,   0.5, 1.2, 1.5, the following are shifted versions, shifted by 0.3,-0.3,.4.
Six items were non-Rasch items generated by randomly permutating the responses (50 data sets, $P=200$). It is seen that the hit rates and the heterogeneity measures are very similar  to the results given in Section \ref{sec:id}.

\subsection{Hierarchical clustering }

\begin{figure}[H]
\centering
\includegraphics[width=7cm]{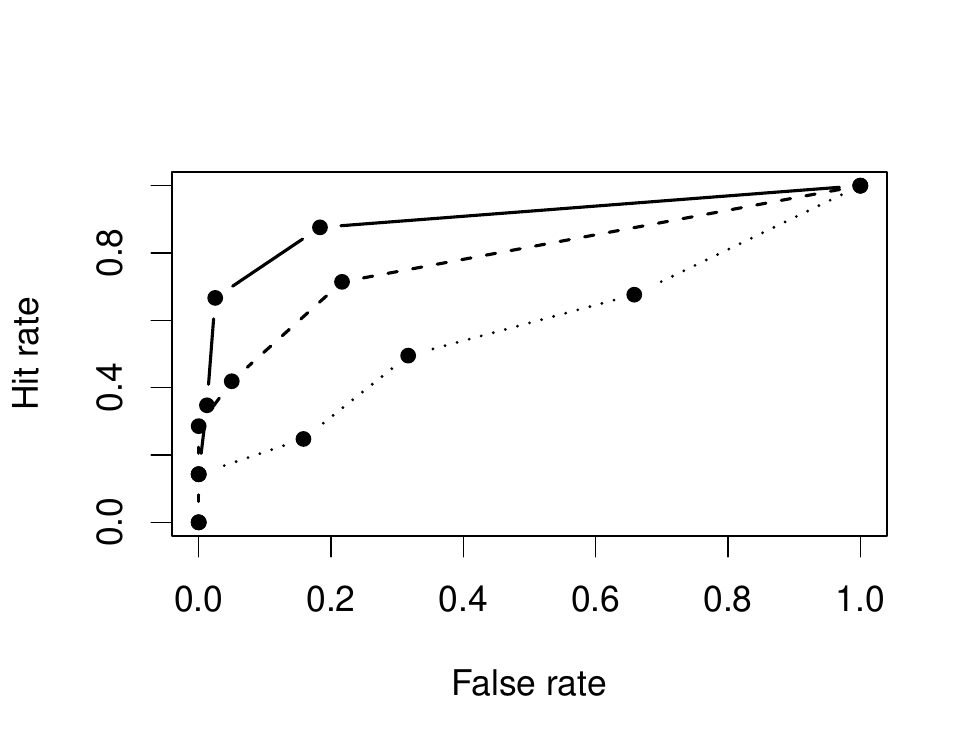}
\includegraphics[width=7cm]{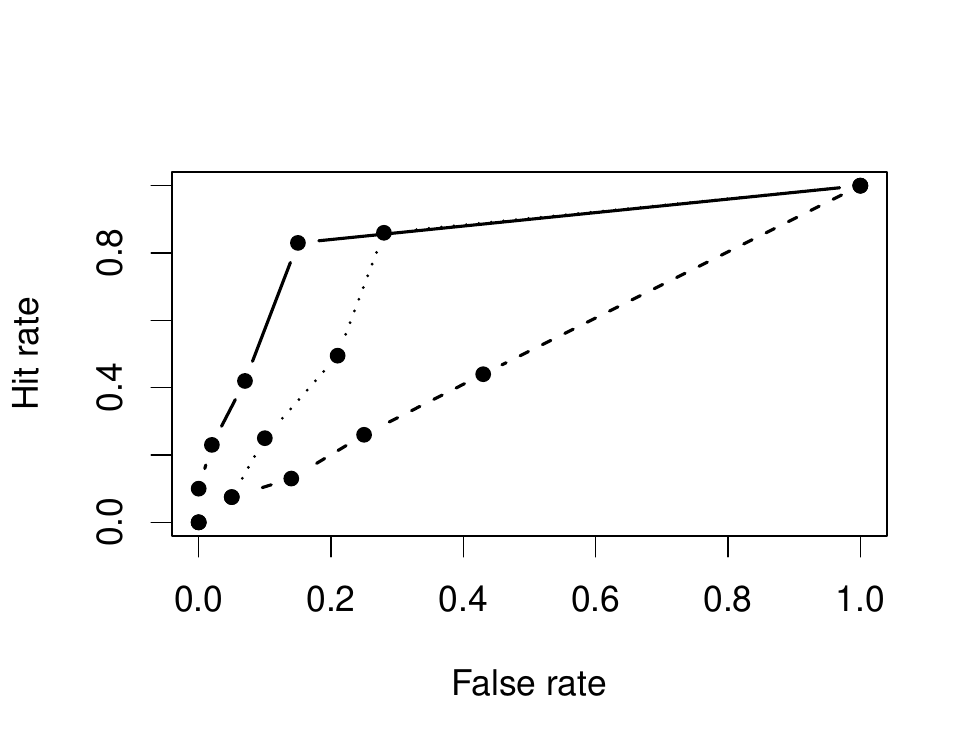}
\caption{Comparison classical clustering (dashed: average linkage, dotted: centroid clustering) for 6 items, true clusters $\{1,\dots,4\}, \{5,6\}$ (first) and   $\{1,\dots,5\}, \{6\}$ ((second));  solid line is proposed cluster procedure ($\sigma_{\theta}=1$, $P=200$)}
\label{fig:comp6varcl}
\end{figure}

Figure \ref{fig:comp6varcl} shows the performance of classical methods and the new method for six items with varying true clusters.
It is again seen that the marginal estimate method performs better than traditional clustering methods.

\end{document}